\documentclass[letterpaper,aps,prd,twocolumn,tightenlines,preprintnumbers,nofootinbib,showkeys,superscriptaddress]{revtex4-1}

 \usepackage{XCharter}
 \usepackage[T1]{fontenc}

\usepackage{bm}
\usepackage{bbold}
\usepackage{pifont}
\usepackage{rotating}
\usepackage{fullpage}
\usepackage{amsfonts}
\usepackage{amsmath}
\usepackage{slashed}
\usepackage{amssymb}
\usepackage{graphicx}
\usepackage{epic}
\usepackage{eepic}
\usepackage{epsfig}
\usepackage{latexsym}
\usepackage[dvipsnames]{xcolor}
\usepackage[export]{adjustbox}
\usepackage{float}
\usepackage{multirow, makecell}
\usepackage{hyperref}
\usepackage{enumitem}
\hypersetup{colorlinks=true,citecolor=red,linkcolor=NavyBlue,urlcolor=NavyBlue}
\usepackage[caption=false]{subfig}
 
\usepackage{natbib}
\usepackage{relsize}
\usepackage[left=1.6cm,right=1.6cm,top=1.9cm,bottom=1.9cm]{geometry}
\usepackage{shorthand}

\usepackage{array}
\newcolumntype{P}[1]{>{\centering\arraybackslash}p{#1}}

\newcommand{\mF}{\mathcal F}

\begin{document}
\relscale{1.05}

\title{Roadmap to explore vectorlike quarks decaying to a new scalar or pseudoscalar}

\author{Akanksha Bhardwaj}
\email{akanksha.bhardwaj@glasgow.ac.uk} 
\affiliation{School of Physics \& Astronomy, University of Glasgow, Glasgow G12 8QQ, UK
}

\author{Tanumoy Mandal}
\email{tanumoy@iisertvm.ac.in}
\affiliation{Indian Institute of Science Education and Research Thiruvananthapuram, Vithura, Kerala, 695 551, India}

\author{Subhadip Mitra}
\email{subhadip.mitra@iiit.ac.in}
\affiliation{Center for Computational Natural Sciences and Bioinformatics, International Institute of Information Technology, Hyderabad 500 032, India}

\author{Cyrin Neeraj}
\email{cyrin.neeraj@research.iiit.ac.in}
\affiliation{Center for Computational Natural Sciences and Bioinformatics, International Institute of Information Technology, Hyderabad 500 032, India}

\date{\today}

\begin{abstract}
\noindent
The current experimental data allow for a sub-TeV colourless weak-singlet scalar or pseudoscalar. If such a singlet field is present together with TeV-range vectorlike top and bottom partners, there is a possibility that the heavy quarks decay dominantly to the singlet state and a third-generation quark, and the singlet state decays to quark and boson pairs. Such a possibility may arise in various models but it has not been explored experimentally, especially in the context of vectorlike-quark searches. We consider some minimal models, covering the possible weak representations of the top and bottom partners, that can be mapped to many well-motivated ultraviolet-complete theories. We chart out the possible interesting and unexplored signatures of the exotic decay of vectorlike quarks and identify benchmark points representing different signal topologies for the high luminosity LHC. We perform a general scan of the parameter space with the relevant direct search bounds and find that large regions, which do not require any fine-tuning, remain open for the unexplored channels. We also perform a simple projection study in the cleanest channel and indicate how other new but experimentally challenging channels can be used to probe more regions of the parameter space.
\end{abstract}

\maketitle

\section{Introduction}
\label{sec:intro}
\noindent
TeV-scale vectorlike quarks (VLQs) are an essential ingredient of many new physics models. Because of their vector-like nature, they do not contribute to the gauge anomalies and are less restricted than their chiral counterparts by the current experimental data. Ones that couple with the third-generation quarks (top and bottom partners, we shall collectively refer to them as top partners) appear in composite Higgs models with a partially-composite top quark~\cite{Kaplan:1983fs,Kaplan:1991dc,Agashe:2004rs,Ferretti:2013kya,Ferretti:2014qta,Ferretti:2016upr}, extra-dimensional models~\cite{Chang:1999nh,Gherghetta:2000qt,Contino:2003ve,Gopalakrishna:2011ef,Gopalakrishna:2013hua,Barcelo:2014kha}, Little-Higgs models~\cite{Arkani-Hamed:2002iiv,Schmaltz:2002wx,Perelstein:2003wd,Martin:2009bg}, etc. However, in the last few years, an extensive direct search program at the LHC has gradually tightened the mass bounds on these quarks. For top partners like the $T$ quark (with electromagnetic charge $2/3$) or the $B$ quark (with charge $-1/3$), the current exclusion limit stands as high as $\sim1.5$ TeV~\cite{CMS:2019eqb,CMS:2020ttz, ATLAS:2021ibc,ATLAS:2021ddx,ATLAS:2022ozf}.

Direct VLQ searches generally assume that they decay exclusively to Standard Model (SM) particles, i.e., to a third-generation quark and a heavy vector-boson or a Higgs. However, this assumption need not hold in general, especially if one looks beyond the minimal models where the top partners can have new decay modes. For example, a top partner can decay to another heavy quark or a new boson~\cite{Gopalakrishna:2013hua,Gopalakrishna:2015wwa,Serra:2015xfa,Anandakrishnan:2015yfa,Banerjee:2016wls,Kraml:2016eti,Dobrescu:2016pda,Aguilar-Saavedra:2017giu,Chala:2017xgc,Moretti:2017qby,Bizot:2018tds,Colucci:2018vxz,Han:2018hcu,Dermisek:2019vkc,Kim:2019oyh,Xie:2019gya,Benbrik:2019zdp,Cacciapaglia:2019zmj,Dermisek:2020gbr,Wang:2020ips,Das:2020ozo,Choudhury:2021nib,Dermisek:2021zjd,Corcella:2021mdl,Dasgupta:2021fzw}. A possibility that has attracted some interest in the current literature is that a vectorlike top partner decays to a new spinless state singlet under the SM gauge group [i.e., $(\mathbf{1},\mathbf{1},0)$ under 
$\mathrm{SU}(3)_c\times\mathrm{SU}(2)_L\times\mathrm{U}(1)_Y$] and a third-generation quark. The singlet state could be a naturally arising pseudoscalar in the non-minimal composite Higgs models~\cite{Serra:2015xfa,Banerjee:2022izw}, a dark matter candidate~\cite{Anandakrishnan:2015yfa}, or just an extra scalar~\cite{Gopalakrishna:2015wwa}.  One can also achieve such a set-up from a bottom-up perspective by extending the Higgs and top sectors of the SM. For example, one can add VLQs in two-Higgs-doublet models~\cite{Benbrik:2019zdp}. There even have been some claims in the literature that the current LHC data points to the existence of a sub-TeV spinless state mainly coupling with the third-generation fermions (see, e.g.,~\cite{vonBuddenbrock:2015ema,vonBuddenbrock:2016rmr,Buddenbrock:2019tua,Richard:2020cav,Arganda:2021yms}).

For the LHC phenomenology of the top-partner models, the addition of a singlet state looks interesting from two perspectives. First, in most well-motivated models (e.g., the ones addressing the gauge hierarchy problem), these quarks are supposed to be not much heavier than a TeV. Hence, the possibility of evading the experimental bounds with a new decay mode are worth looking into. In other words, instead of giving up on the models, the strong bounds can be taken as a motivation for considering next-to-minimal (but otherwise well-motivated) models with TeV-scale particles. Second, a new lighter-than-TeV singlet (pseudo)scalar allowed by the current data leads to a host of new possibilities to probe the top partners. In this paper, we attempt to quantify the above points and sketch a roadmap for how these possibilities can be explored at the LHC.

We consider some simple phenomenological models based on the possible weak representations of the top partners to describe their interactions with a singlet scalar $\phi$ or a pseudoscalar $\eta$. With these generic models, we recast the latest LHC limits on $T$ and $B$ to see how low the limits can go with the increasing branching ratio (BR) in the extra modes and how that affects the model parameters. For simplicity,  we assume $\Ph=\{\phi,\et\}$ has exclusive couplings only with the top partners initially. However, since the heavy quarks mix with their SM partners after Electroweak symmetry breaking (EWSB), $\Phi$ can decay to $qq$ final states at the tree level. Moreover, depending on its mass, it can also decay to gauge boson pairs through quark loops. Even though $\Phi$ has no direct couplings with the light quarks, it can still be produced directly at the LHC through the loop-mediated $gg\to\Phi$ process. This allows us to put limits on the $\Phi gg$ coupling. Taking these (and the other applicable) bounds into account, we perform a parameter scan on these generic models to show that there is no need to fine-tune the parameters to satisfy the bounds. Since our parametrisation is generic, the parameters easily relate to a broad class of complete models. We chart out the interesting signal topologies that the new decay mode could lead to and, for a fixed heavy-quark mass scale, present a sample set of benchmark points leading to different decay topologies. We also explain the intuitions behind the choice of parameter points so that one can choose a similar set of parameters for a different heavy-quark mass.

The plan of the paper is as follows. In the next section, we present the generic parametrisation of the phenomenological models. In Sec.~\ref{sec:decays}, we work out the parameter relations and the decays, in Sec.~\ref{sec:scan} we present the bounds and parameter scans, in Sec.~\ref{sec:projection} we discuss the possible new topologies, the benchmark set and a simple projection study for the $T$ in the $t\gm\gm$ mode at the high-luminosity LHC (HL-LHC). Finally, we conclude  in Sec~\ref{sec:conclusion}.

\section{Phenomenological models}\label{sec:model}
\noindent
In general, top partners can have various weak representations. However, since the singlet $\Phi$ is present in our case, the top partners must be weak-singlets or form a weak-doublet to make their interaction with $\Phi$ and the SM quarks gauge invariant. Therefore, we consider two types of models: one with one weak-singlet vectorlike top partner (either $T$ or $B$) and a $\Phi$, and the other with a weak-doublet of $T$ and $B$ and a $\Phi$. We look into the possibilities separately. We assume the weak singlet $\Phi$ does not acquire a vacuum expectation value (VEV) and has no direct mixing with the SM Higgs. 

\subsection{Singlet VLQ}

\noindent
Before EWSB, the terms contributing to the masses of the weak-singlets $T$ and $B$ (that transform as $(\mathbf{3},\mathbf{1},2/3)$ and $(\mathbf{3},\mathbf{1}, -1/3)$, respectively) can be parametrised as,
%
\begin{align}
\label{eq:Lagsingvlqgen}
\mc{L} \supset&  - \Big\{\tilde{\lm}_{q}\lt(\bar{Q}_LH_F\rt)q_R +\om_{F}\lt(\bar{Q}_LH_F\rt)F^\prime_R\nonumber\\
&+ \tilde{\om}_{F} m_F \bar{F}^\prime_Lq_R 
+ M_{F}\bar{F}^\prime_LF^\prime_R+ {\tr h.c.}\Big\},
\end{align}
where $F$ is either $T$ or $B$ with $q$ denoting the corresponding third-generation quark, $H_{\{T,B\}}  = \{\widetilde H,H\} = \{i\sigma^2H^*,H\}$, with $H$ being the SM Higgs doublet, $Q_L$ is the third-generation quark doublet, $M_F$ is the  VLQ mass scale, and $\tilde{\lm}_{q}$, $\omega_F$ and $\tilde{\omega}_F$ are all dimensionless couplings. In general, the SM gauge symmetry allows the off-diagonal mixing term, $\tilde{\om}_{F} m_F\bar{F}^\prime_Lq_R+$ h.c, between fields with the same quantum numbers. In the underlying theory, such a term can come from a high-scale symmetry breaking, or from a finite overlap in the bulk wavefunctions of the two fields in the extra-dimensional theories, etc. However, this is a redundant degree of freedom as one can always absorb this term with a simple  redefinition:
\begin{align}
\label{eq:transformation}
F^\prime_L \to F_L , \quad
F^\prime_R \to F_R - \dfrac{\tilde{\om}_{F} m_F}{M_F}q_R.
\end{align}
With the above replacements, the additional  $\bar{F}^\prime_Lq_R$ term disappears and, in the new basis, the previous Lagrangian looks as,
\begin{align}
\label{eq:Lagpheno}
\mc{L} \supset -\Big\{&\lm_{q}\lt(\bar{Q}_LH_F\rt)q_R
+\om_{F}\lt(\bar{Q}_LH_F\rt)F_R\nn\\
& + M_{F}\bar{F}_LF_R+ {\tr h.c.}\Big\},
\end{align}
where $\lm_q$ is now the redefined Yukawa coupling, 
%
\begin{align}
\lm_q = \tilde{\lm}_q - \omega_F\tilde{\omega}_F\dfrac{m_F}{M_F}.
\end{align}
Hence, we get the following mass matrix after EWSB,  
\begin{align}
\mc{L}_{mass}^F =& \bpm \bar{q}_L & \bar{F}_L \epm 
\bpm 
\begin{array}{cc}
\lm_q \frac v{\sqrt2} & \l\om_{F}\,\frac v{\sqrt2} \\ 0 & M_F
\end{array} 
\epm 
\bpm q_R \\ F_R \epm + {\tr h.c.},\label{eq:massmats}
\end{align}
where $v$ the Higgs VEV. The interactions between $\Ph$ and $F$ can be written as,
\begin{align}
\mc{L}_{int}^{\Phi F} =& - \lm_{\Phi F}^a\Phi\,\bar{F}_L\Gm F_R - \lm_{\Phi F}^b\Phi\,\bar{F}_L\Gm q_R   + {\tr h.c.}
\end{align}
where $\Gm=\lt\{1,i\gm_5\rt\}$ for $\Ph =\lt\{\ph,\et\rt\}$.

\subsection{Doublet VLQ}
\noi
When $T$ and $B$ together forms a weak-doublet, $\mF=\left(T\ B\right)^T=(\mathbf{3},\mathbf{2},1/6)$,
we can write the terms relevant for the quark masses after eliminating the redundant doublet-doublet off-diagonal mixing term ($\sim m_F\bar\mF_R Q_L$) as,
\begin{align}
\mc{L} \supset&- \Big\{\lm_{t}\lt(\bar{Q}_L\widetilde{H}\rt)t_R + \rh_{T}\lt(\bar{\mF}_L \widetilde{H}\rt)t_R  + \lm_{b}\lt(\bar{Q}_LH\rt)b_R\nn\\
&\ 
+\rh_{B}\lt(\bar{\mF}_L H\rt)b_R + M_{F}\bar\mF_L\mF_R+ {\tr h.c.}\Big\}.
\end{align}
From this, we get the following mass matrices,
\begin{align}
\mc{L}_{mass}^\mF =& \bpm \bar{t}_L & \bar{T}_L \epm 
\bpm 
\begin{array}{cc}
\lm_t \frac v{\sqrt2}  & 0 \\ \rh_{T}\,\frac v{\sqrt2} & M_T
\end{array} 
\epm 
\bpm t_R \\ T_R \epm\nn\\
& +\bpm \bar{b}_L & \bar{B}_L \epm 
\bpm 
\begin{array}{cc}
\lm_b\frac v{\sqrt2}  & 0 \\ \rh_{B}\,\frac v{\sqrt2} & M_B
\end{array} 
\epm 
\bpm b_R \\ B_R \epm + {\tr h.c.}\label{eq:massmatd}
\end{align}
The interactions between $\Ph$ and the doublet $\mF$ can be written as,
\begin{align}
\mc{L}_{int}^{\Phi \mF} =& - \lm_{\Phi D}^a\Phi\,\bar{\mF}_L\Gm\mF_R - \lm_{\Phi D}^b \Phi\,\bar{\mF}_R\Gm Q_L   + {\tr h.c.}
\end{align}

\section{Mass eigenstates and decays}\label{sec:decays}
\noi
The mass matrices in Eqs.~\eqref{eq:massmats} and \eqref{eq:massmatd} can be diagonalised by the following  bi-orthogonal rotations,
\begin{align}
\bpm t_P \\ T_P \epm =& \bpm c^T_P & s^T_P \\ -s^T_P & c^T_P \epm \bpm {t_1}_P \\ {t_2}_P \epm, \\
\bpm b_P \\ B_P \epm =& \bpm c^B_P & s^B_P \\ -s^B_P & c^B_P \epm \bpm {b_1}_P \\ {b_2}_P \epm,
\end{align}
where $P=\{L,R\}$ is the chiral projection, $\{c^F_P, s^F_P\}= \left\{\cos\theta_{F_P},\sin\theta_{F_P}\right\}$ and $\{q_1,q_2\}$ are the mass eigenstates.
If we generically express the mass matrix for $F$ as
\begin{align}
\mc M = \left(\begin{array}{cc}
m_{q}  & \m_{F1} \\ \m_{F2} & M_F
\end{array}\right), \label{eq:massmatrix}
\end{align}
we can express the left and right mixing angles as
\begin{align}
\tan{(2\theta_{F_L})} =& \frac{2\lt(m_q\,\m_{F2}+M_F\,\m_{F1}\rt)}{\lt(m_q^2+\m_{F1}^2\rt)-\lt(M_F^2+\m_{F2}^2\rt)},\label{eq:mixangleL}\\
\tan{(2\theta_{F_R})} =& \frac{2\lt(m_q\,\m_{F1}+M_F\,\m_{F2}\rt)}{\lt(m_q^2+\m_{F2}^2\rt)-\lt(M_F^2+\m_{F1}^2\rt)} .\label{eq:mixangleR}
\end{align}
The mass eigenvalues $m_{q_1,q_2}$ are given by
\begin{align}
m_{q_1,q_2}^2 = \frac12\Bigg[&{\rm Tr}\lt(\mc M^{\rm T}\mc M\rt) \nn\\ 
&\mp\sqrt{\lt[{\rm Tr}\lt(\mc M^{\rm T}\mc M\rt)\rt]^2-4\lt({\rm Det}~\mc M\rt)^2}\Bigg].\label{eq:massevs}
\end{align}
We identify $q_1$ with the physical SM quark. The above expressions indicate for a very heavy $F$, i.e., when $M_F\gg m_q, \m_{F1}, \m_{F2}$, the SM-quark and the VLQ effectively decouple.

\subsection{Decays of the VLQs}
\noi
There are two new particles, $t_2$ and $\ph$ (or $\et$), in the spectrum of the singlet $T$ model. Due to  $t\leftrightarrow T$ mixing, the $t_2$ quark can decay to $Wb$, $Zt$ and $ht$ final states (from here on, we drop the subscripts from $t_1$ and $b_1$ and simply refer to them as $t$ and $b$, respectively). Moreover, the $t_2$ quark can also decay to $\phi t$ (or $\et t$) mode if $M_\Ph + M_t <M_{t_2}$. We list the interactions responsible for these decays.
\begin{itemize}[leftmargin=*]
\item Interactions with the gauge bosons ($W$ and $Z$):
\begin{align}
\mc{L} \supset &\frac{g}{\sqrt{2}}s_L\,\bar{b}_L\gm^\mu t_{2L} W_\mu^{-}
+ \frac{2g \mathbb T^t_3}{\cos\theta_W} c_Ls_L\,\bar{t}_{L}\gm^\mu t_{2L} Z_\mu +{\tr h.c.}
\label{eq:LagintWZ}
\end{align}
where $\mathbb T^t_3 =1/2$ is the weak-isospin of $t_L$. We drop the superscripts from $c^T_L$ and $s^T_L$ when their meaning is clear from the context.

\item Interactions with the Higgs boson ($h$):
\begin{align}
\mc{L} \supset \frac{1}{v}\bigg[&\lt(m_t\,c_Ls_R + \m_{T1}\,c_Lc_R\rt)\bar{t}_{L}t_{2R}\nn\\
& + \lt(m_t\,s_Lc_R - \m_{T1}\,s_Ls_R\rt)\bar{t}_{R}t_{2L}\bigg]h + {\tr h.c.}
\end{align}

\item Interactions with $\phi$ (or $\eta$):
\begin{align}
\mc{L} \supset & - \lm_{\Ph T}^a\Ph \lt( c_L\bar{t}_{2L}-s_L\bar{t}_{L}\rt)\Gamma\lt(c_R {t}_{2R}-s_R {t}_{R}\rt) \nn\\
                      & -\lm_{\Ph T}^b \Ph \lt(c_L\bar{t}_{2L} - s_L\bar{t}_{L}\rt)\Gamma\lt(c_R {t}_{R} + s_R {t}_{2R}\rt) + {\tr h.c.}\label{eq:LagintS}
\end{align}
\end{itemize}

In the singlet $B$ model, $b_2$ can decay to $Wt$, $Zb$, $hb$ and $\Phi b$ final states. The interaction terms responsible for the decay of $b_2$ can be obtained from Eqs.~\eqref{eq:LagintWZ}-\eqref{eq:LagintS} by  $\{t,t_2\}\leftrightarrow \{b,b_2\}$. The only exception is the interaction with the $Z$ boson, which picks up a minus sign since $\mathbb T^b_3 =-1/2$.

In the doublet model, the gauge interactions of the VLQs responsible for their decays are given as,
\begin{align}
\mc{L} \supset &\ \frac{g}{\sqrt{2}}\Bigg[\lt(c^B_L s^T_L - c^T_L s^B_L\rt)\left(\bar{b}_{L}\gm^\mu t_{2L} W_\mu^{-}-\bar{t}_{L}\gm^\mu b_{2L} W_\mu^{+}\right)\nn\\
& +\Big(s^B_L s^T_L\,\bar{t}_{2L}\gm^\mu b_{2L} + \sum_{X=L,R}c^T_X c^B_X\,\bar{t}_{2X}\gm^\mu b_{2X}\Big)W_\mu^{+}\Bigg]\nn\\ 
& - \frac{2g}{\cos\theta_W}\Big(\mathbb T^T_3\ c^T_R s^T_R\, \bar t_R\gm^\mu t_{2R}\nn\\
&\hspace{1.75cm}+\mathbb T^B_3\ c^B_R s^B_R\, \bar b_R\gm^\mu  b_{2R}\Big)Z_\mu +{\tr h.c.}
\end{align}
where $\mathbb T^T_3=-\mathbb T^B_3 =1/2$.
The interactions with the Higgs boson are given as,
\begin{align}
\mc{L} \supset \frac{1}{v}\bigg[&\lt(m_t\, c^T_L s^T_R - \m_{T2}\,s^T_L s^T_R\rt)\bar{t}_{L}t_{2R}\nn\\
 + &\lt(m_t\, c^T_R s^T_L + \m_{T2}\,c^T_L c^T_R\rt)\bar{t}_{R}t_{2L}\nn\\
 + &\lt(m_b\, c^B_L s^B_R - \m_{B2}\,s^B_L s^B_R\rt)\bar{b}_{L}b_{2R}\nn\\
 + &\lt(m_b\, c^B_R s^B_L + \m_{B2}\,c^B_L c^B_R\rt)\bar{b}_{R}b_{2L}\bigg]h + {\tr h.c.}
\end{align}
and the interactions with $\Ph$ are given as,
\begin{align}
\mc{L} \supset & -\sum_{q=t,b}\Big[\lm_{\Ph D}^a\Ph \lt( c_L\bar{q}_{2L}-s_L\bar{q}_{L}\rt)\Gamma\lt( c_R\bar{q}_{2R}-s_R\bar{q}_{R}\rt) \nn\\
                      & +\lm_{\Ph D}^b \Ph \lt(c_R\bar{q}_{2R} - s_R\bar{q}_{R}\rt)\Gamma\lt(c_L\bar{q}_{L} + s_L\bar{q}_{2L}\rt)\Big] \nn\\
                      &+ {\tr h.c.}\label{eq:LagintD}
\end{align}

\begin{figure*}
\captionsetup[subfigure]{labelformat=empty}
\subfloat[\quad\quad(a)]{\includegraphics[width=0.8\columnwidth]{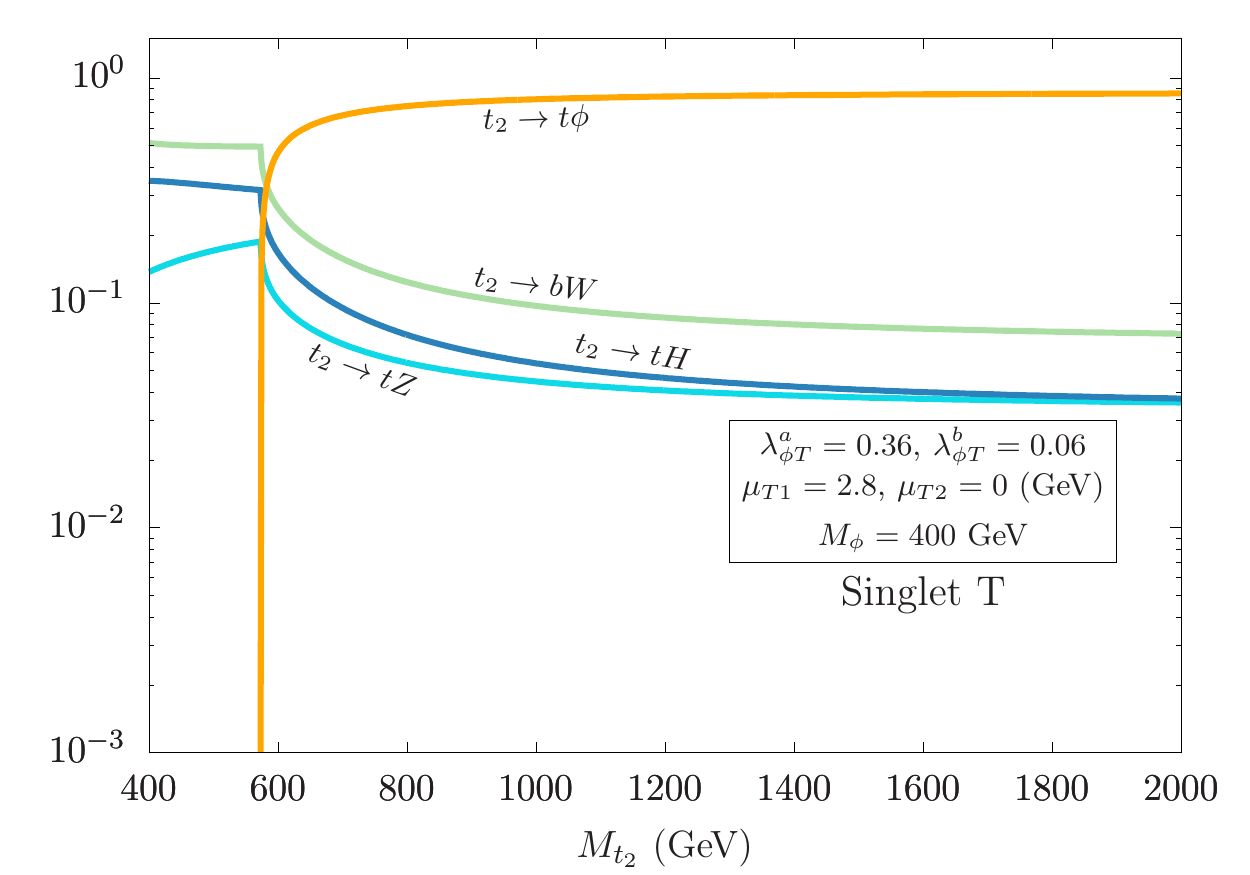}\label{fig:br_TSingT}}\hspace{0.5cm}
\subfloat[\quad\quad(b)]{\includegraphics[width=0.8\columnwidth]{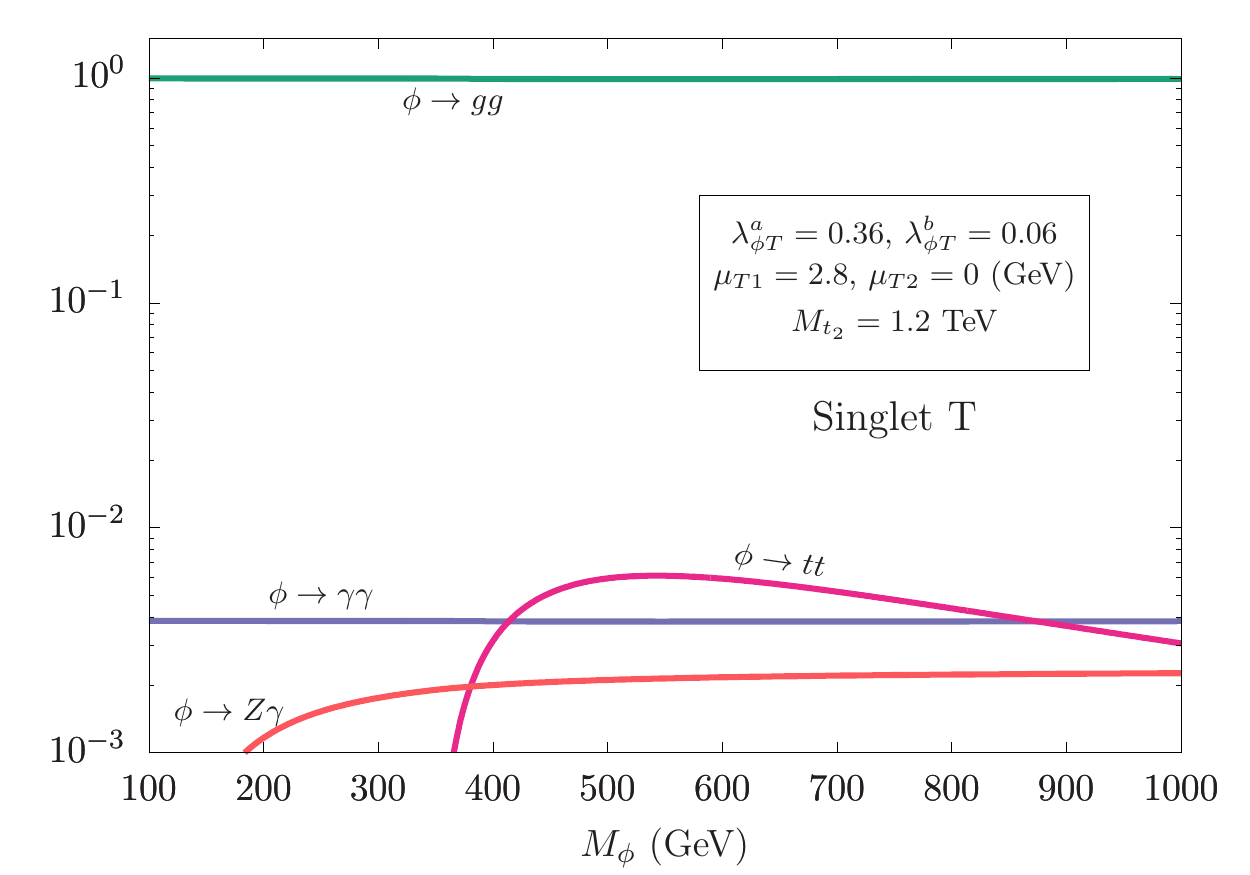}\label{fig:br_TSingPhi}}\\
\subfloat[\quad\quad(c)]{\includegraphics[width=0.8\columnwidth]{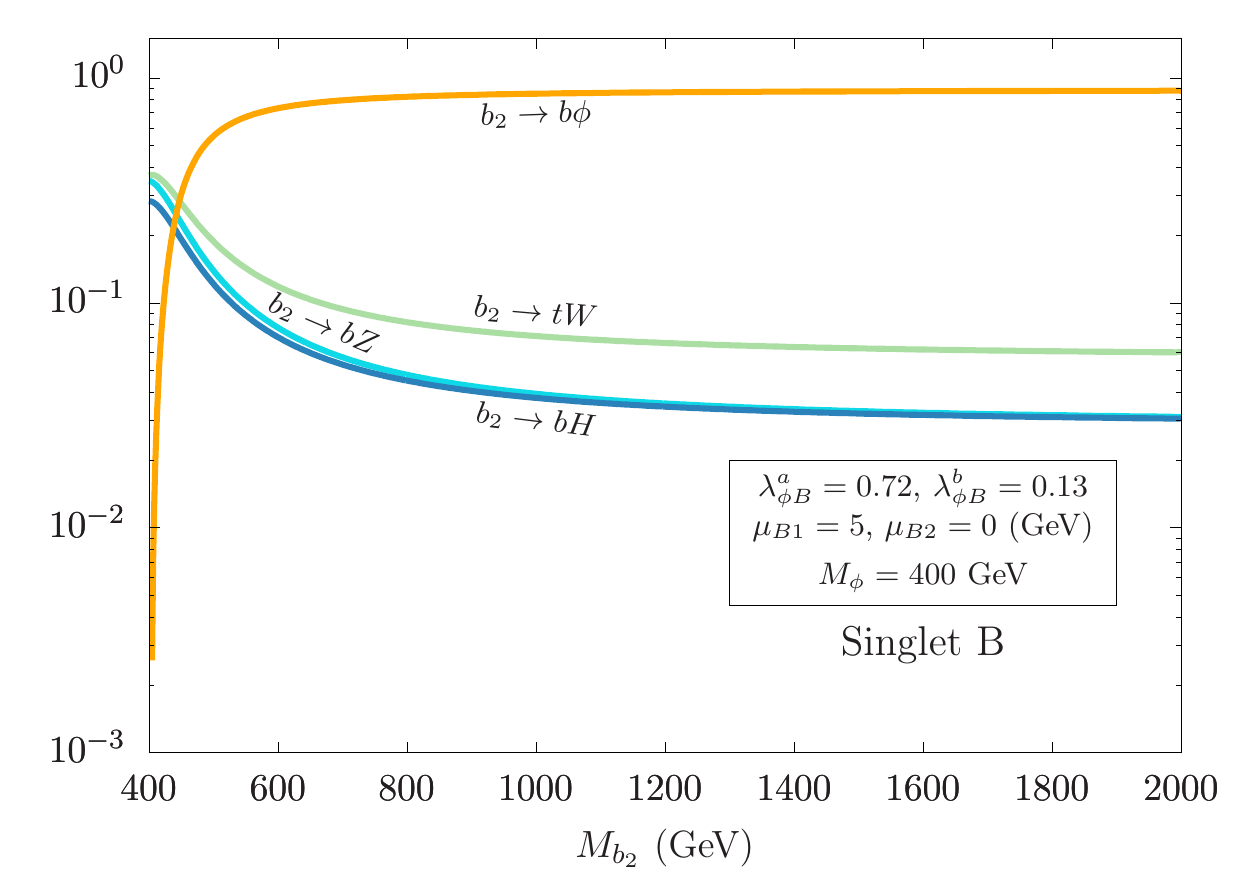}\label{fig:br_BSingB}}\hspace{0.5cm}
\subfloat[\quad\quad(d)]{\includegraphics[width=0.8\columnwidth]{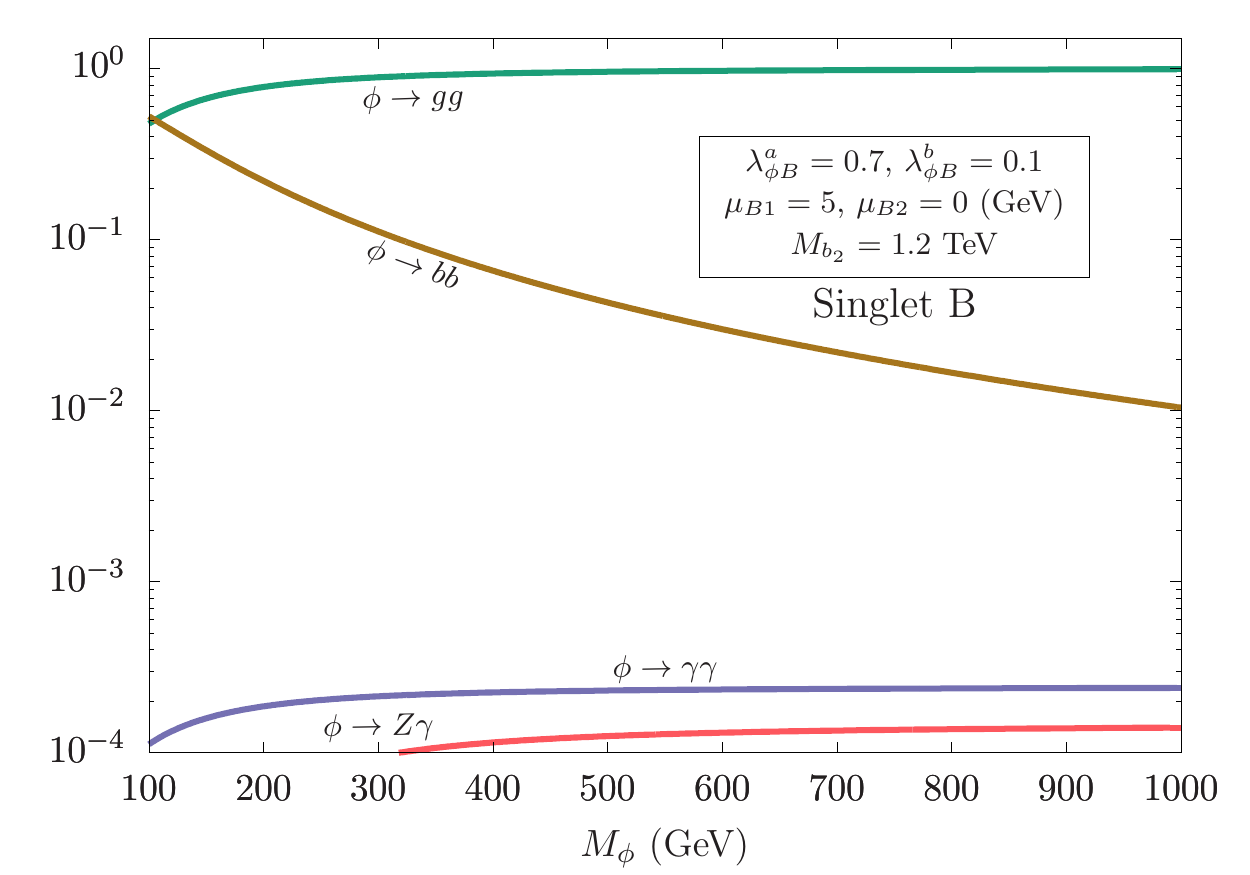}\label{fig:br_BSingPhi}}\\
\subfloat[\quad\quad(e)]{\includegraphics[width=0.8\columnwidth]{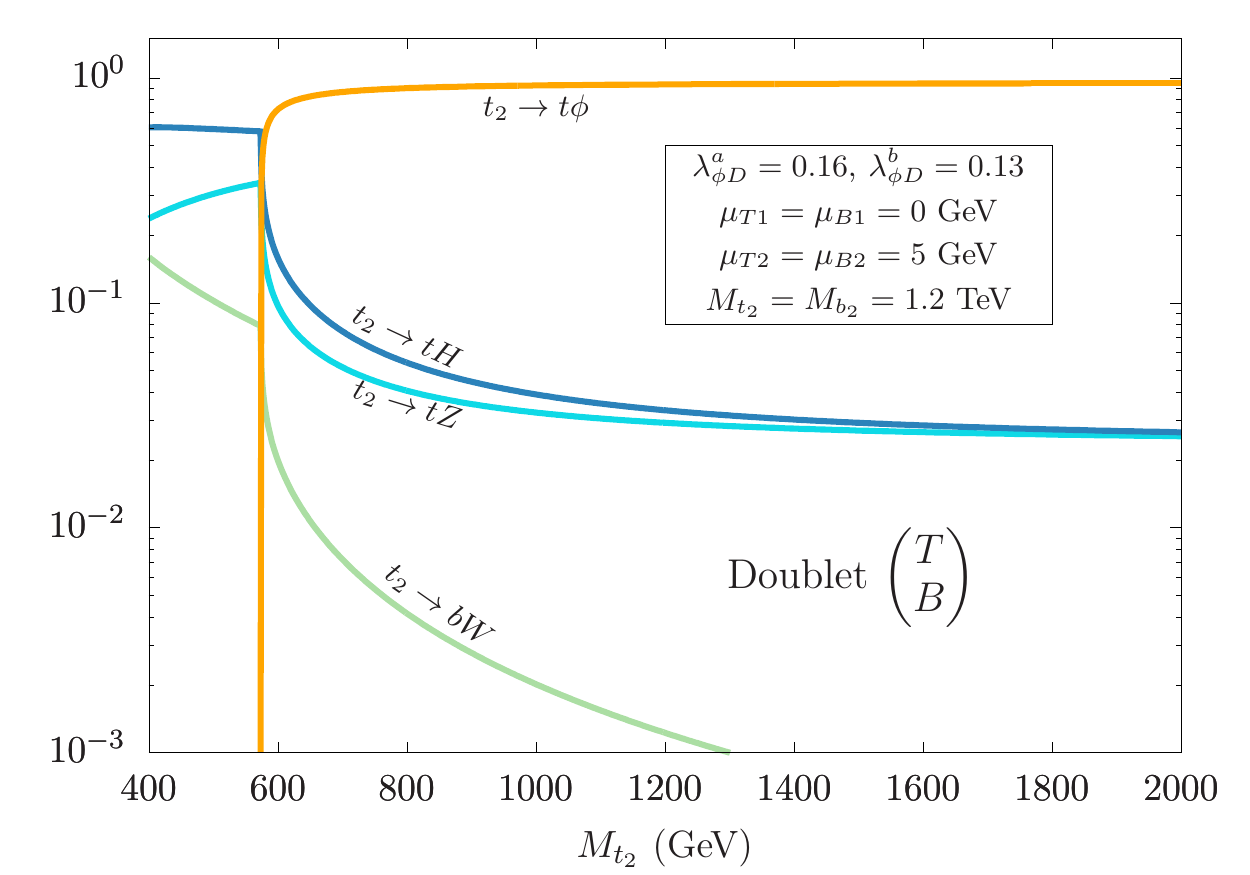}\label{fig:br_DoubT}}\hspace{0.5cm}
\subfloat[\quad\quad(f)]{\includegraphics[width=0.8\columnwidth]{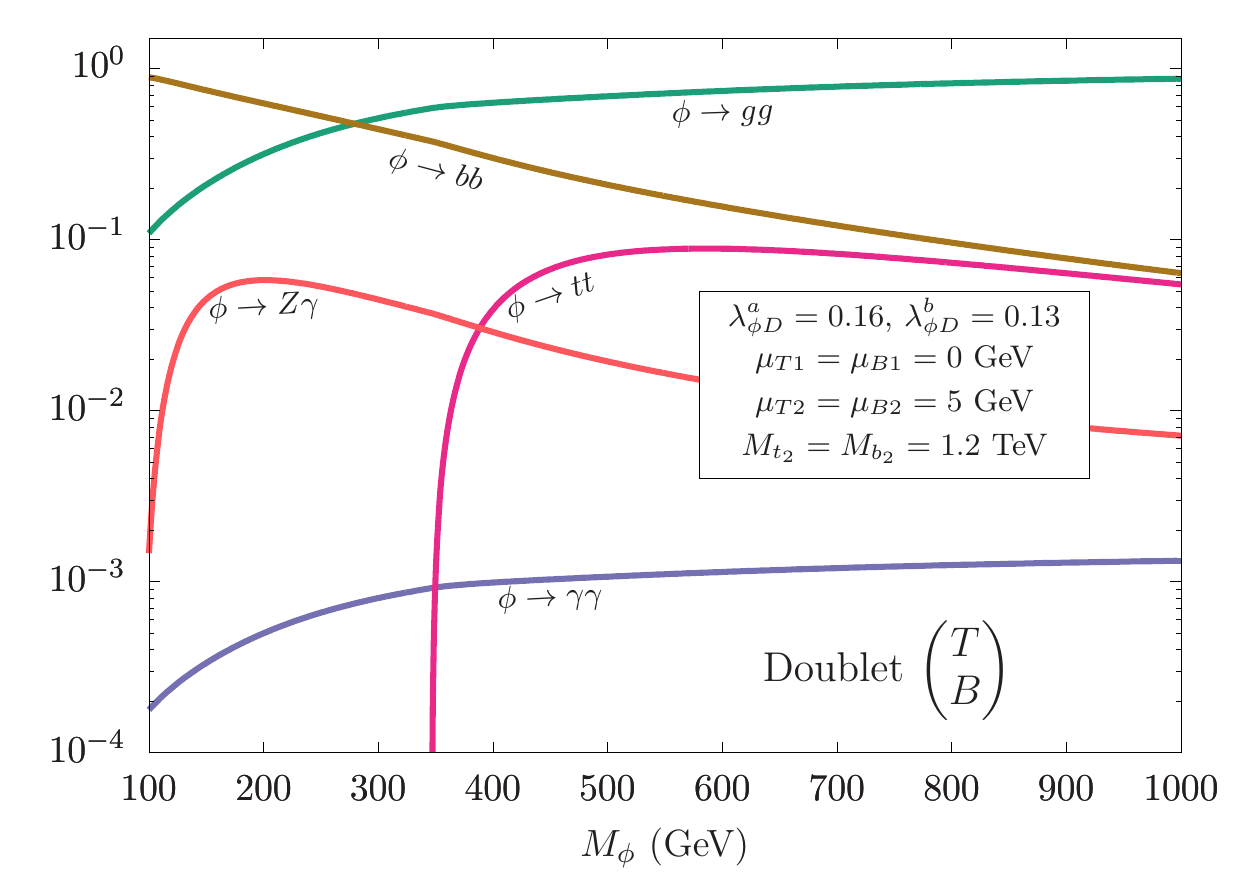}\label{fig:br_DoubPhi}}\\\hspace{1.65cm}
\subfloat[\quad\quad(g)]{\includegraphics[width=0.8\columnwidth]{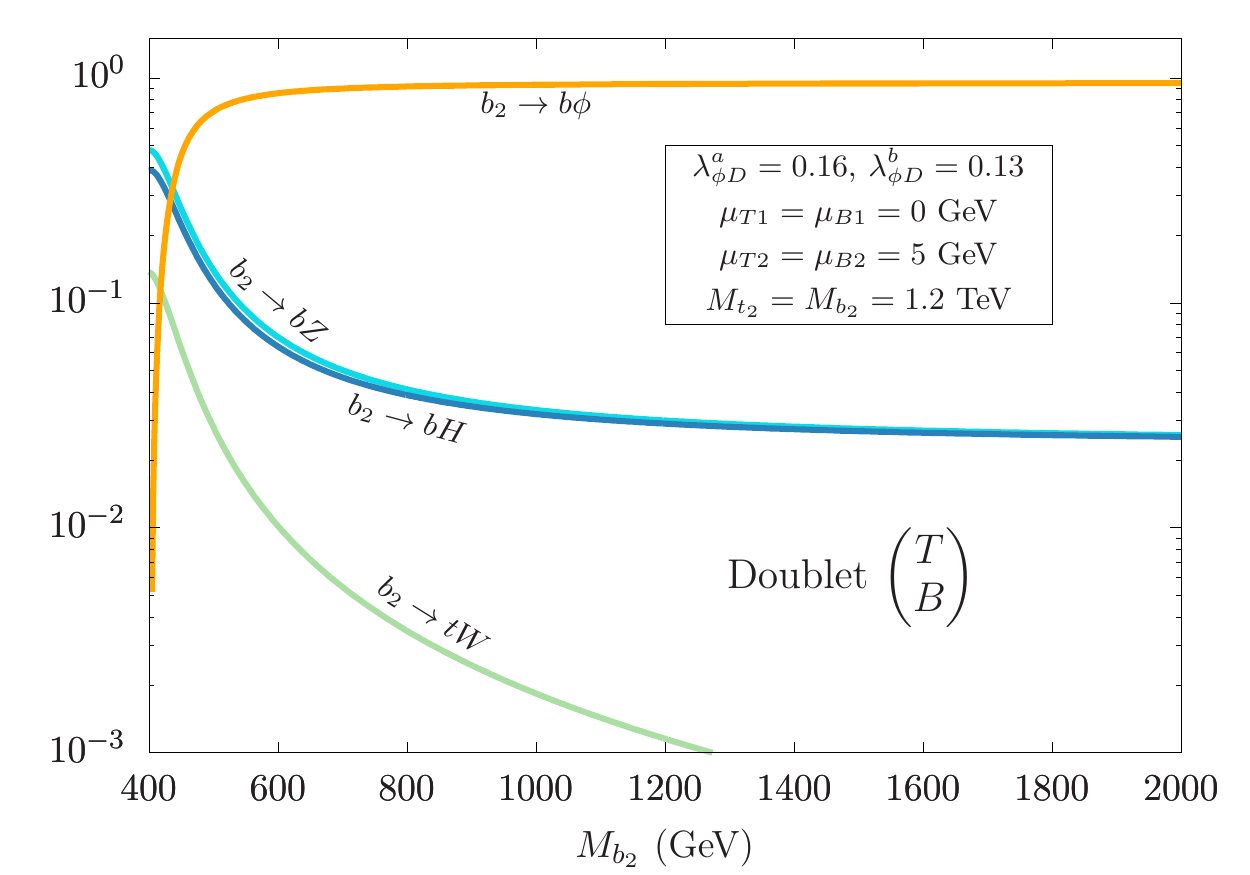}\label{fig:br_DoubB}}\hfill~
\caption{Branching ratio plots for $t_2$, $b_2$ and $\phi$ in ((a), (b)) the singlet T model, ((c), (d)) the singlet B model, and  ((e), (f),  (g)) the doublet model. The set of parameters for each plot is chosen such that $q_2 \to q_1\phi$ is the dominant decay mode for a TeV $q_2$. }
\label{fig:br}
\end{figure*}

\subsection{Additional decays and generic parametrisation}
\noi
In the singlet VLQ  models, we can generically parametrise the terms relevant for the $q_2$ decay as
\begin{align}
\mc{L} \supset&\ C^V_L~\bar{q}^V_{1L}\gm^{\mu}q_{2L}V_{\mu} 
+ C^V_R~\bar{q}^V_{1R}\gm^{\mu}q_{2R}V_{\mu}\nn\\
&+ C^S_L~\bar{q}_{1R}q_{2L}S + C^S_R~\bar{q}_{1L}q_{2R}S + {\tr h.c.}\label{eq:genericlag}
\end{align}
where $V=\{Z,W\}$, $q^{\left\{Z,W\right\}}_{1}=\{q_1,q_1^\prime\}$, and $S=\{h,\ph,\et\}$. We can express the partial decay widths of $q_2$ as
\begin{align}
\Gamma_{q_2\to q^V_1V} = 
&\Big[	\Big\{(C^V_L)^2 + (C^V_R)^2\Big\}
\Big\{\left(1 - x_{q_1^V}^2\right)^2- 2x_V^4\nn\\
& + x_V^2\left(1 + x_{q_1^V}^2\right) \Big\}- 12C^V_L C^V_R\  x_{q_1^V}x_V^2 \Big]\nn\\
&\times \frac{\mc P(M_{q_2},x_{q_1^V},x_{V})}{x^2_{V}}
\label{Gamq2q1V.EQ} \\
\Gamma_{q_2\to q_1S} = 
&\Big[\left\{(C^S_L)^2 + (C^S_R)^2\right\}
\left\{\left(1 - x_{q_1}^2 - x_S^2\right)^2 \right\}\nn\\
& + 4C^S_L C^S_R\ x_{q_1}\Big]\times \mc P(M_{q_2},x_{q_1},x_{S}),
\label{Gamq2q1h.EQ}
\end{align}
where $x_{q_1} \equiv M_{q_1}/M_{q_2}$, $x_V \equiv M_V/M_{q_2}$, $x_S \equiv M_S/M_{q_2}$, and
\begin{align}
\mc P(M,x,y) \equiv \frac{M}{32\pi^2}\sqrt{1+x^4+y^4-2x^2-2y^2 - 2x^2y^2 }. \nn
\end{align}
For $M_\Ph< 2M_{q_2}$, the neutral spinless particle $\Ph$ mainly decays to $gg$, $\gm\gm$, $Z\gm$, $ZZ$ and 
$q_1q_1$ final states. The decay to $q_2q_2$ is kinematically forbidden and a singlet $\Phi$ cannot decay to $WW$ mode.
The decays to the vector bosons are mediated through $q_1$ or $q_2$ loops. The terms in the Lagrangian responsible for the decay of $\Ph$ can be expressed as,
\begin{align}
\mc{L} \supset \sum_{i=1,2}C^i_\Ph \left(\bar{q}_{iL}\Gm q_{iR}+\bar{q}_{iR}\Gm q_{iL}\right)\Ph.
\end{align}
The partial width for the tree-level $\Ph\to q_1q_1$ decay is given by
\begin{align}
\Gm_{\Phi\to q_1q_1} = \frac{3\left(C^1_\Phi\right)^2M_\Phi}{8\pi}\lt(1-\frac{4m_{q_1}^2}{M_\Phi^2}\rt)^{3/2}
\end{align}
and the partial widths of the loop-induced decay channels are given as~\cite{Gunion:1989we,Fontes:2014xva} (also see~\cite{Djouadi:2005gi,Djouadi:2005gj}),
\begin{align}
\Gm_{\Ph\to\gm\gm} =&\ \frac{\al^2M_\Ph^3N_{c}^2}{256\pi^3}\lt|\sum_{i=1,2} \frac{C^i_\Ph Q_{q_i}^2 }{m_{q_i}}F_{1/2}^\Ph\left(\frac{4m_{q_i}^2}{M_\Ph^2}\right)\rt|^2,\label{eq:S2gmgm}\\
\Gm_{\Ph\to gg} =&\ \frac{\al_{s}^2M_\Ph^3}{128\pi^3}\lt|\sum_{i=1,2} \frac{C^i_\Ph }{m_{q_i}}F_{1/2}^\Ph\left(\frac{4m_{q_i}^2}{M_\Ph^2}\right)\rt|^2,\label{eq:S2gg}
\end{align}
\begin{align}
\Gm_{\Ph\to Z\gm}= &\ \frac{\al^2M_\Ph^3N_{c}^2}
{32\pi^3\sin^2\theta_w\cos^2\theta_w}
\lt(1-\frac{M_Z^2}{M_\Phi^2}\rt)^3\nn\\&\times\lt|\sum_{i=1,2} \frac{-Q_{q_i}g^i_VC^i_\Ph  }{m_{q_i}} I_\Ph\left(\frac{4m_{q_i}^2}{M_\Ph^2},\frac{4m_{q_i}^2}{M_Z^2}\right)\rt|^2,
\end{align}
where $g^i_V$ is the vector projection of the $Z\bar q_i q_i$ coupling:
\begin{align}
&\left.\begin{array}{ll}
g^1_V =&  c^2_L \mathbb T^{q}_3 -2 Q_{q_i} \sin^2\theta_W\\~\\
g^2_V =&  s^2_L \mathbb T^{q}_3 -2 Q_{q_i} \sin^2\theta_W
\end{array}\right\}({\rm singlet}~Q), \\
&\left.\begin{array}{ll}
g^1_V =&  (1+s_R^2) \mathbb T^{q}_3 -2 Q_{q_i} \sin^2\theta_W\\~\\
g^2_V =&  (2-s_R^2)\mathbb T^{q}_3 -2 Q_{q_i} \sin^2\theta_W
\end{array}\right\}({\rm doublet}).
\end{align}
The loop functions $F_{1/2}^\Ph(\ta)$ are well known:
\begin{align}
F_{1/2}^\phi(\tau) =&\ 2\tau\lt[1+\lt(1-\tau\rt)f(\tau)\rt],\nn\\
F_{1/2}^\eta(\ta) =&\ 2\tau f(\tau)
\end{align}
with
\begin{align}
f(\tau) =&\ \theta(\tau -1)\lt[\sin^{-1}\lt(\frac{1}{\sqrt{\tau}}\rt)\rt]^2\nn\\
&\ - \theta(1-\tau)\frac{1}{4}\lt[\ln\lt(\frac{1+\sqrt{1-\tau}}{1-\sqrt{1-\tau}}-i\pi\rt)\rt]^2.
\end{align}
The function $I_\Phi(\ta,\lm)$ is defined as,\footnote{The loop functions, $I_\phi(\ta,\lm)=I_{1}(\ta,\lm)-I_{2}(\ta,\lm)$ and $I_\eta(\ta,\lm)=I_{2}(\ta,\lm)$ in the notation of Ref.~\cite{Gunion:1989we}.}
\begin{align}
I_\phi(\ta,\lm) =&\ \frac{\ta\lm}{2(\ta-\lm)}
+\frac{\ta^2\lm^2+\ta\lm(\ta-\lm)}{2(\ta-\lm)^2}\lt[f(\ta)-f(\lm)\rt]\nn\\
&\ +\frac{\ta^2\lm}{(\ta-\lm)^2}\lt[g(\ta)-g(\lm)\rt]
\end{align}
with
\begin{align}
g(\ta) =&\ \theta(\tau -1)\lt[\sqrt{\ta-1}\sin^{-1}\lt(\frac{1}{\sqrt{\tau}}\rt)\rt]\nn\\
&\ + \theta(1-\tau)\left[\frac{\sqrt{1-\ta}}{2}\ln\lt(\frac{1+\sqrt{1-\tau}}{1-\sqrt{1-\tau}}-i\pi\rt)\right]
\end{align}
and
\begin{align}
I_\eta(\ta,\lm) =&\ -\frac{\ta\lm}{2(\ta-\lm)}\lt[f(\ta)-f(\lm)\rt].
\end{align}
The above expressions are valid in the doublet VLQ model as well, if one assumes $q_1=\{t,b\}$, $q_2=\{t_2,b_2\}$ and lets the summations run over all the four quarks. Generally, the scalar
$\phi$ would also decay to massive vector bosons like $ZZ$ or $WW$ (in the doublet model). However, since these decays are smaller than the $Z\gamma$ decay, we ignore their contribution to the $\phi$ total width. In principle, in the models with two heavy quarks, $\Ph$ can also decay to the lighter of the two heavy quarks if it is kinematically allowed. 

We illustrate the decays of the heavy particles  in Fig.~\ref{fig:br} for a representative  choice of parameters in the three models. In the plots, we only show the scalar $\phi$, but one gets similar plots for $\eta$.  In the following sections, we show that the model parameter space that yields high branching for the new decay channels of the heavy quarks does not require any fine tuning, i.e., it is quite open.

\section{Constraints \& the available parameter space}\label{sec:scan}
\noindent 
We are interested in the parameter region(s) where the $q_2 \to q_1\Phi$ decay is significant. Eqs.~\eqref{eq:LagintS} and ~\eqref{eq:LagintD} indicate that it is possible to get high branching ratio (BR) for $q_2 \to q_1\Phi$ decay channel for a sizeable $\lambda^{a,b}_{\Phi q_2}$ or $\lm^{a,b}_{\Phi D}$. However a large $\lambda^{a,b}_{\Phi q_2}$ or $\lambda^{a,b}_{\Phi D}$ would enhance the $gg\to\Phi$ production at the LHC and hence, might be constrained. We look into the bounds on $\Ph$ parameters and the mass bounds on the heavy quarks for the LHC limits on these models.  If $M_\Phi > 2 m_q$, $\Phi$ decays to $q\bar q$ states. This mode often becomes significant (note that the coupling of $\Phi$ to $q$ is suppressed by the sine(s) of the mixing angle(s)). 
However, of all the decay modes of $\Ph$, the di-photon mode offers one of the cleanest signals in the entire mass range of $\Phi$ in all models~\cite{Mandal:2016bhl,Danielsson:2019ftq,Danielsson:2016nyy}; we use it to put bounds on $\Phi$ parameters. 

\begin{figure*}
\captionsetup[subfigure]{labelformat=empty}
\centering
\subfloat[\quad\quad(a)]{\includegraphics[width=\columnwidth]{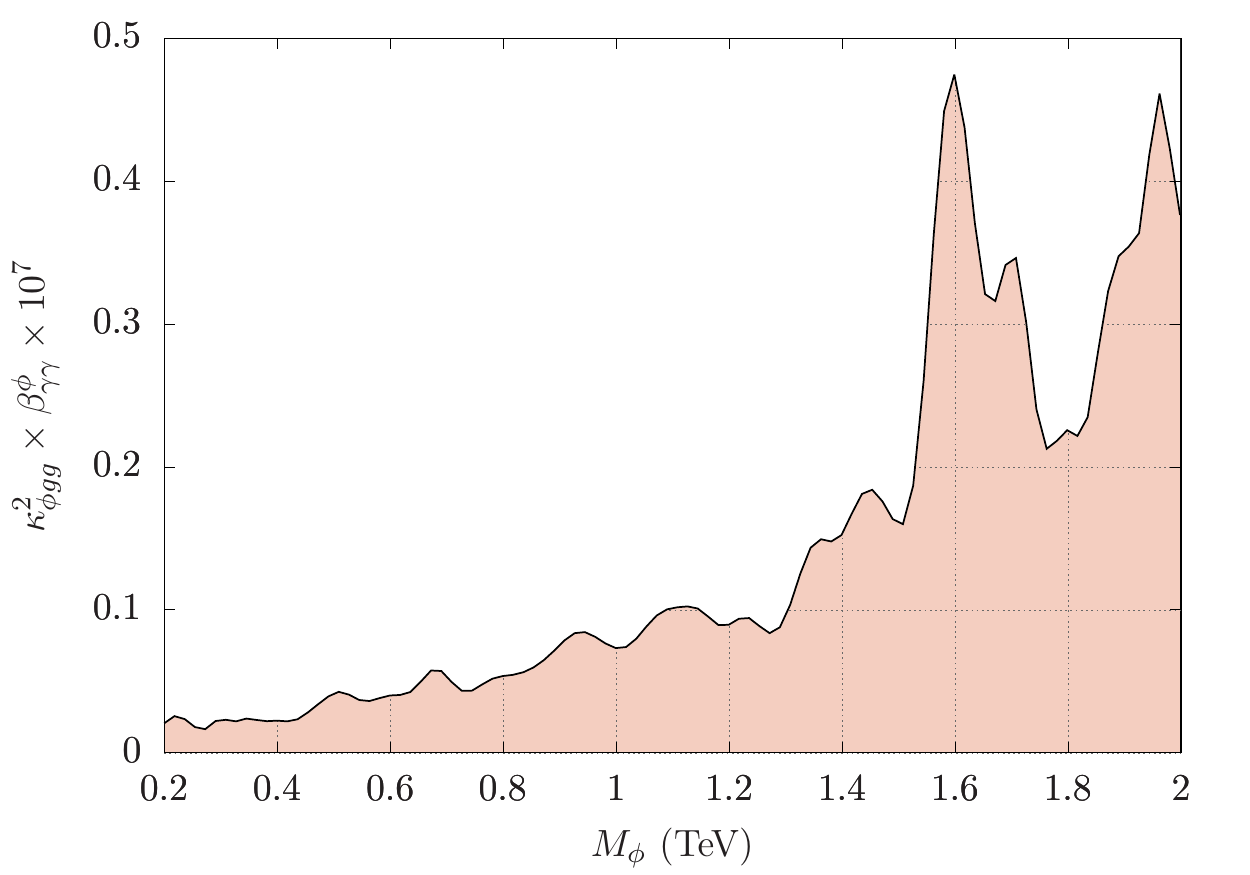}\label{fig:boundsphi}}\hfill
\subfloat[\quad\quad(b)]{\includegraphics[width=\columnwidth]{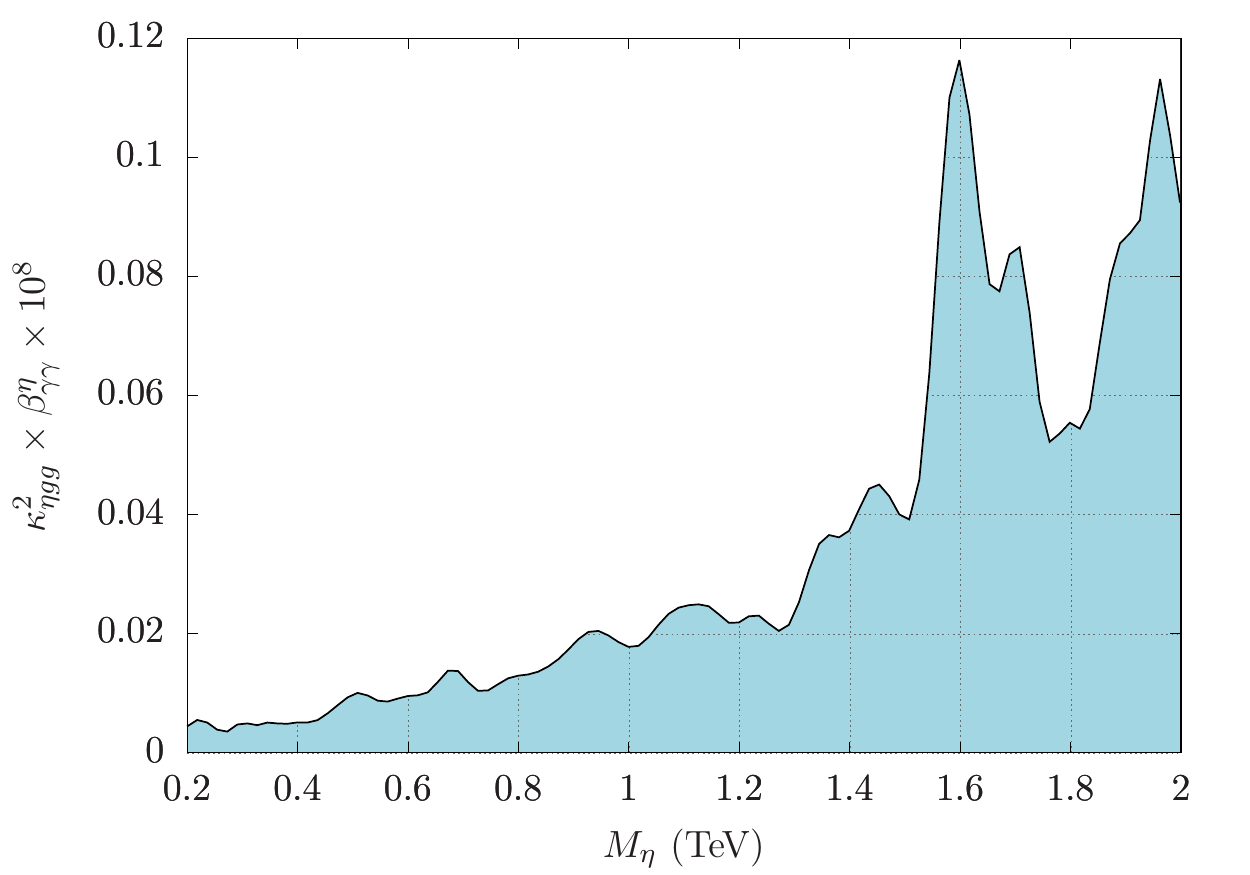}\label{fig:boundseta}}
\caption{Limits on the square of the (a) scalar and (b) pseudoscalar couplings with a pair of gluons times the di-photon branching ratio  from the LHC data~\cite{ATLAS:2021uiz}. The white regions are excluded.}
\label{fig:boundsPhi}
\end{figure*}
\begin{figure*}
\captionsetup[subfigure]{labelformat=empty}
\centering
\subfloat[\quad\quad(a)]{\includegraphics[width=\columnwidth]{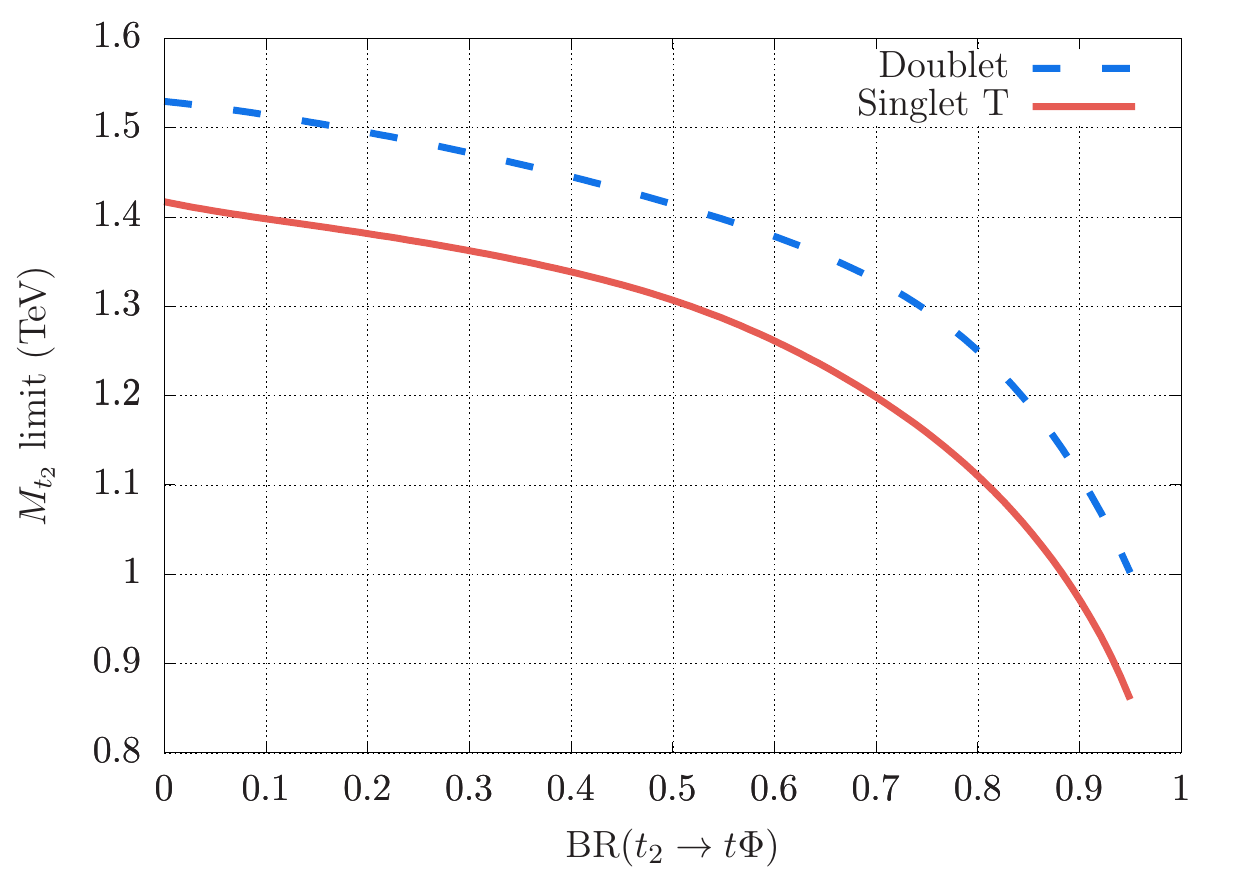}\label{fig:lims_B2}}\hfill
\subfloat[\quad\quad(b)]{\includegraphics[width=\columnwidth]{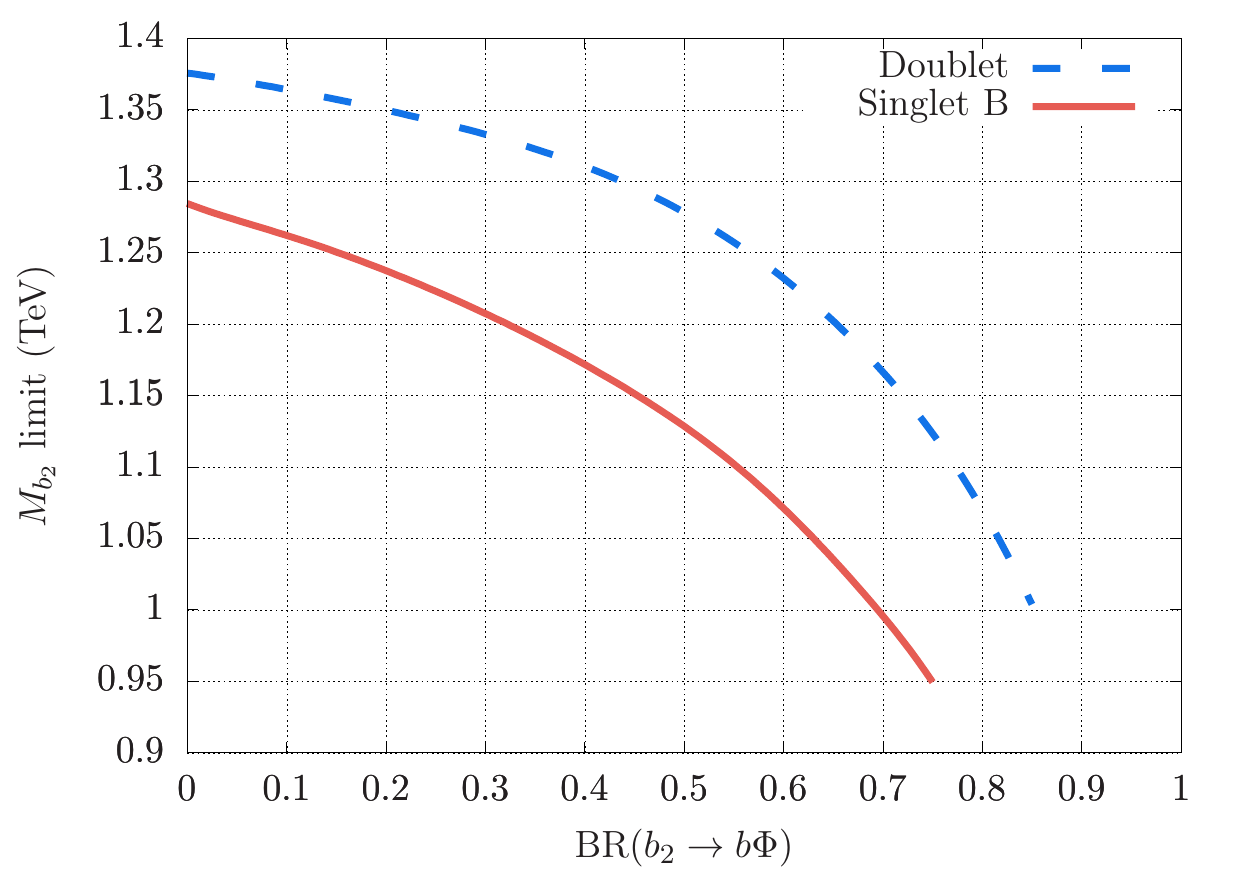}\label{fig:lims_T2}}
\caption{LHC exclusion limits on (a) $t_2$ and (b) $b_2$ in the singlet and doublet models as functions of the branching ratio in the extra decay mode. }
\label{fig:bounds_vlQ_rescaled}
\end{figure*}

So far, there have been several searches for heavy resonances decaying to two photons from the ATLAS~\cite{ATLAS:2017ayi,ATLAS:2021uiz} and CMS~\cite{CMS:2016xbb,CMS:2018dqv} collaborations. We use the latest ATLAS search data~\cite{ATLAS:2021uiz} to recast the bounds. Since we are interested in top partners heavier than $\Phi$, we can consider the following  $5$-dimensional effective Lagrangian to model the direct production of $\Phi$ at the LHC, 
\begin{equation}
     \mathcal{L} \supset \frac{g^2_s}{v} \left[ \kappa_{\phi g g} \phi G^{a}_{\mu \nu} G^{a\; \mu \nu} + \kappa_{\eta g g} \eta G^{a}_{\mu \nu} \tilde{G}^{a\; \mu \nu} \right].
     \label{sca_pSca_L}
\end{equation}
The above Lagrangian lets us recast the di-photon bound on the fiducial production cross section times BR$(X\to\gm\gm)$ in terms of $\kappa^2_{\Phi gg}\times \bt^\Phi_{\gm\gm}$, where $\bt^\Phi_{\gm\gm}$ is the BR of $\Phi$ in the $\gm\gm$ mode:
\begin{align}
\kappa^2_{\Phi gg} \times \bt^\Phi_{\gm\gm}    <\lt. \frac{\sg_{\rm fid}\times {\rm BR~}(X\to\gm\gm)}{\ep\times K_\Phi\times 
\sg_{\rm LO}(gg\to\Phi)}\rt|_{\kappa_{\Phi gg}=1}.
\end{align}
Here, $\ep$ is the reported efficiency, $K_\Phi$ is the NNLO QCD $K$-factor, which we take as the same as for the Higgs, $K_\Phi  \approx \sg_{\rm NNLO}^h/\sg_{\rm LO}^h \approx 2.5$~\cite{Spira:1995mt}. We show the recast limits in Fig.~\ref{fig:boundsPhi}. The generic parametrisation of the $\Phi$ couplings allows easy interpretation of the limits in terms of diagonal couplings of the quarks with $\Phi$. However, the off-diagonal couplings, i.e., $\lambda^{a}_{\Phi F}$ and  $\lambda^{b}_{\Phi F}$ or  $\lm^{a}_{\Phi D}$ and $\lm^{b}_{\Phi D}$ are unrestricted by these limits.

Similarly, both the ATLAS and CMS collaborations have been actively looking for the VLQs as well. The direct LHC searches for $T$ and $B$ assume they can only decay to a third-generation quark and an SM gauge or Higgs boson. With the introduction of $\Phi$, this assumption breaks down and we get,
\begin{equation}
     \beta_{q_1^\prime W} + \beta_{q_1Z} + \beta_{q_1h} = 1 - \beta_{q_1\Phi},
     \label{br_constraint}
\end{equation}
where $\beta_{f_i}$ is the BR for the $q_2 \to f_i$ decay. For $M_{q_2}\gtrsim$ TeV, $\beta_{q_1^\prime W} \approx 2\beta_{q_1Z} \approx 2\beta_{q_1h}$ in the singlet models and $ \beta_{q_1Z} \approx \beta_{q_1h},\ \beta_{q_1^\prime W}\approx 0$ in the doublet models.
One can obtain the new mass exclusion limits from the exclusive pair production searches (often presented for $100\%$ BR in one of the SM decay modes) by rescaling the theory cross section lines with the square of the corresponding BR. Similarly, it is also possible to calculate the exclusion limits from the inclusive searches. Assuming the inclusive signal selection efficiencies remain unaffected by the presence of an additional decay mode (a reasonable assumption given the inclusive nature of the signals), $\sigma^{incl}\left(pp\to \bar q_2q_2\to f_i+X\right)$ scales with a factor,
\begin{align}
\mc B_{f_i}^{incl} = \beta_{f_i}^2  + 2\sum_{j\neq i}\beta_{f_i}\beta_{f_j}= \beta_{f_i}\left(2 - \beta_{f_i}\right),
\end{align}
where the factor $2$ in the middle comes from combinatorics.
For a value of $\bt_{q_1\Phi}$, we first recast the relevant limits from the available  exclusive~\cite{CMS:2020ttz, CMS:2019eqb} and inclusive~\cite{ATLAS:2021ibc} searches to select the strongest one. We show the new limits on $M_{q_2}$ in Fig.~\ref{fig:bounds_vlQ_rescaled}. With increasing  $\bt_{q_1\Phi}$, the limits on the heavy quarks relax significantly. 

There are searches for single production of the singlet top partners by the ATLAS and CMS collaborations~\cite{ATLAS:2022ozf,ATLAS:2021ddx}. However, unlike the pair production, single productions are model dependent, i.e., their cross sections depend on unknown coupling(s). As a result, the exclusion limits from single production searches depend not only on the BRs but also on the absolute magnitude of the unknown VLQ coupling parametrised as $\kappa_T$ in Ref.~\cite{Buchkremer:2013bha}. If $\kappa_T$ is of order one, the single-production search limits on VLQs become stronger than the pair-production limits. However, for small off-diagonal mass matrix elements ($\mu_{F_i}/M_F \lesssim 0.1$), $\kappa_T$ becomes small ($< \lm_{QED}$) making the single production limits weaker than  the pair production ones. For this study, we stick to the regions of the parameter space where $\kappa_T$ is much smaller than unity by restricting the range of the off-diagonal elements of the mass matrix, i.e. ($\mu_{F_i} \lesssim 50$ for $M_F\sim$ TeV). 

Apart from the direct search limits, there are limits on the $Z$ boson coupling to the left-handed $b$-quark, i.e., $\kp_{Z\bar b_Lb_L}$ in the models with a $B$ quark since the coupling shifts from its SM value due to $b$-$B$ mixing (unless some symmetry prevents it--see e.g. Refs.~\cite{Agashe:2006at,Gopalakrishna:2013hua}). The measurements of $R_b$ and $\Gamma_b$ at LEP~\cite{ParticleDataGroup:2020ssz} restrict 
 $\Delta\kp_{Z\bar b_Lb_L}$ to be less than about $1\%$ (roughly, $\left(1-c^B_L\right)^2=\left(s^B_L\right)^2\lesssim0.1)$. The  direct limits from flavour-changing neutral couplings~\cite{Nir:1990yq} also restrict the mixing parameters between the SM quarks and their partners from being arbitrarily large.

\begin{figure*}
\captionsetup[subfigure]{labelformat=empty}
\subfloat[\quad\quad(a)]{\includegraphics[width=0.48\columnwidth]{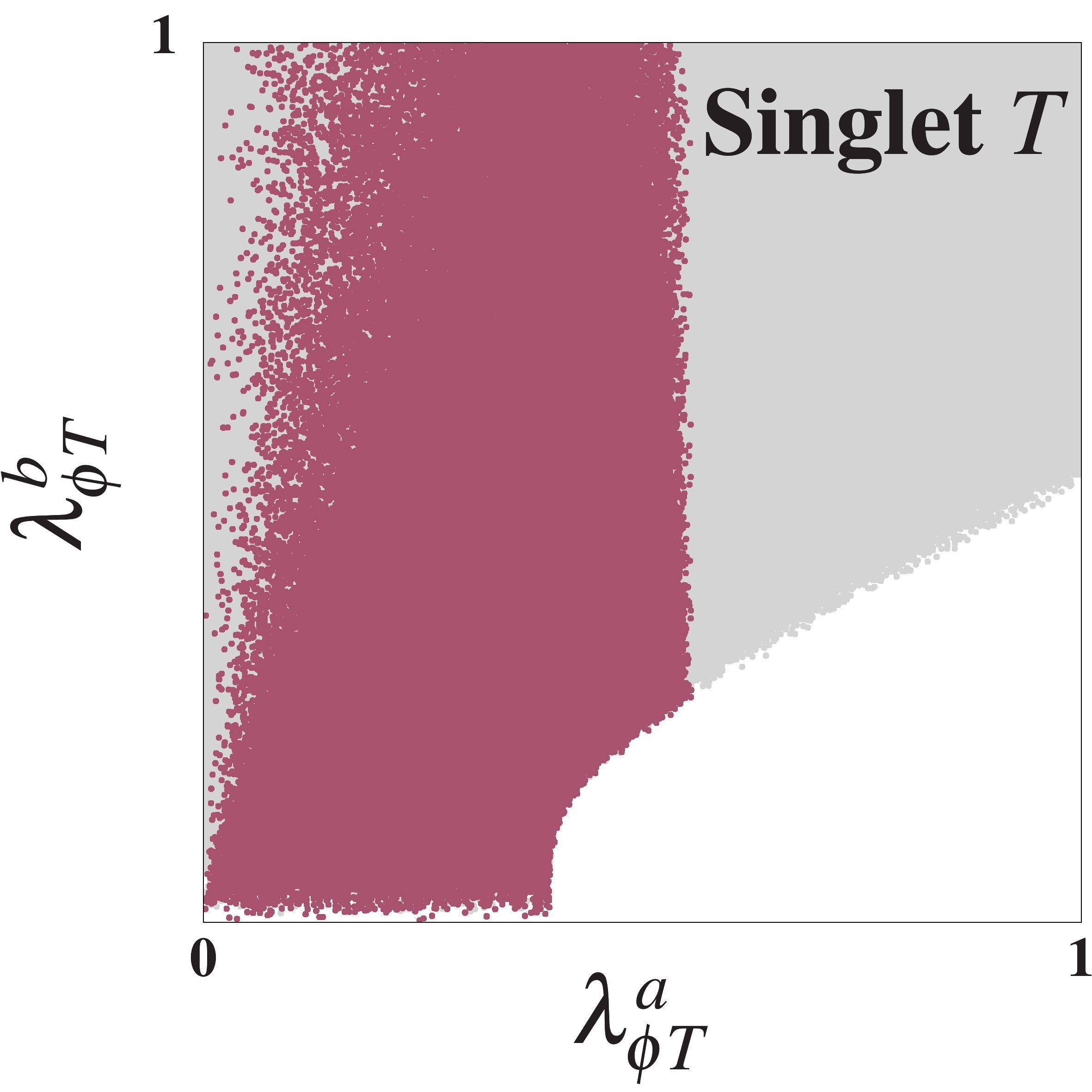}\label{fig:phi_SingTlm}}\hspace{0.2cm}
\subfloat[\quad\quad(b)]{\includegraphics[width=0.48\columnwidth]{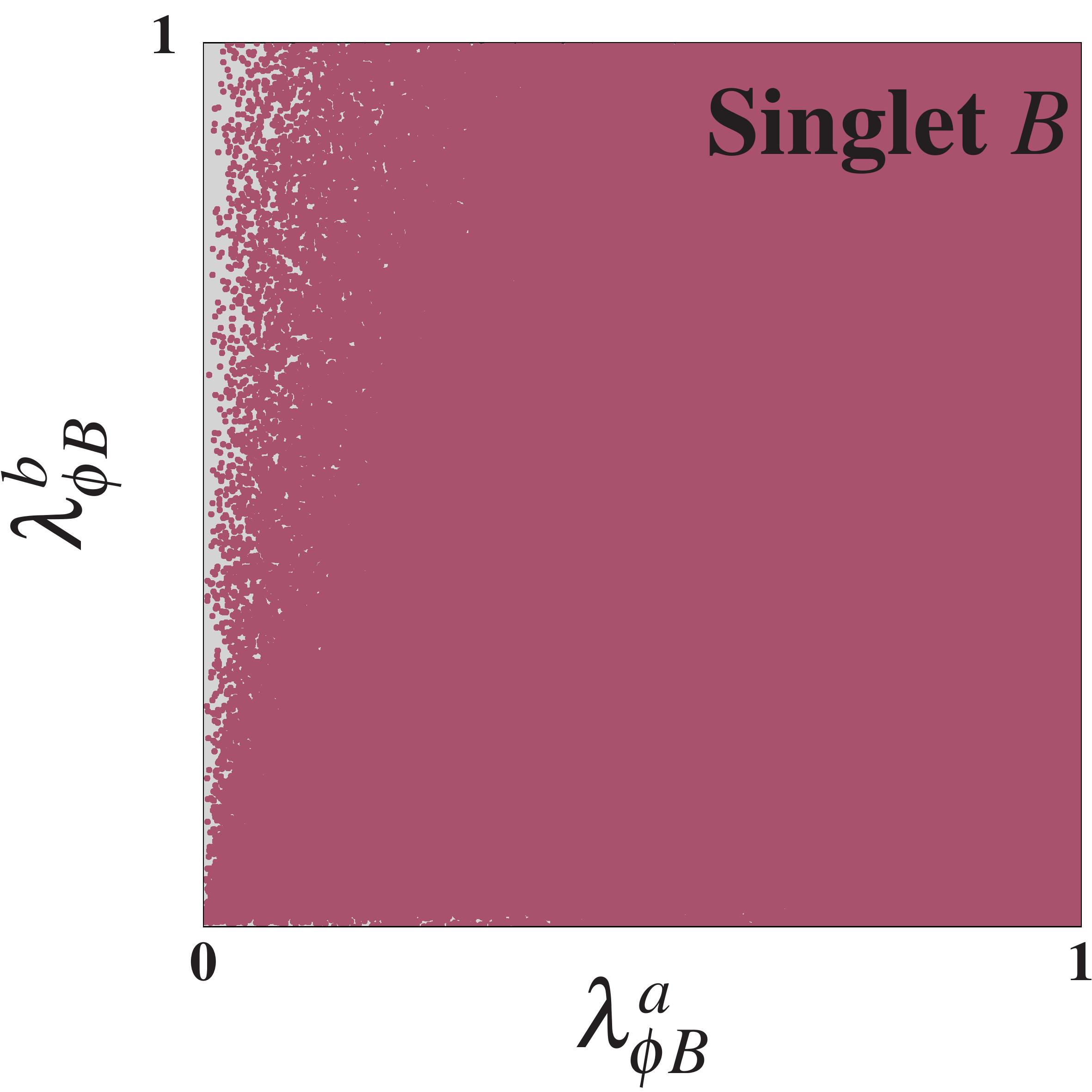}\label{fig:phi_SingBlm}}\hspace{0.2cm} 
\subfloat[\quad\quad(c)]{\includegraphics[width=0.48\columnwidth]{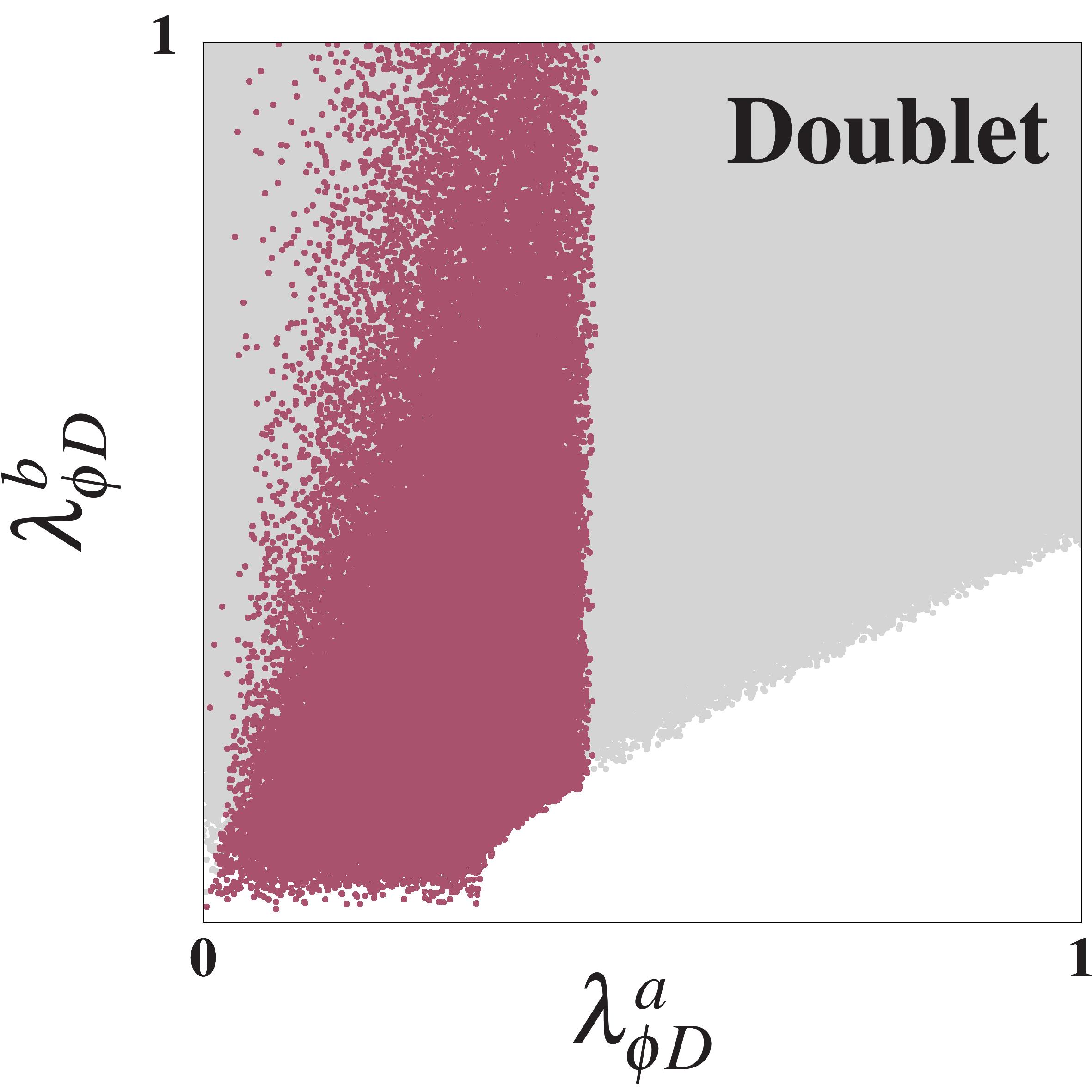}\label{fig:phi_Doublm}}\hspace{0.2cm} 
\subfloat[\quad\quad(d)]{\includegraphics[width=0.48\columnwidth]{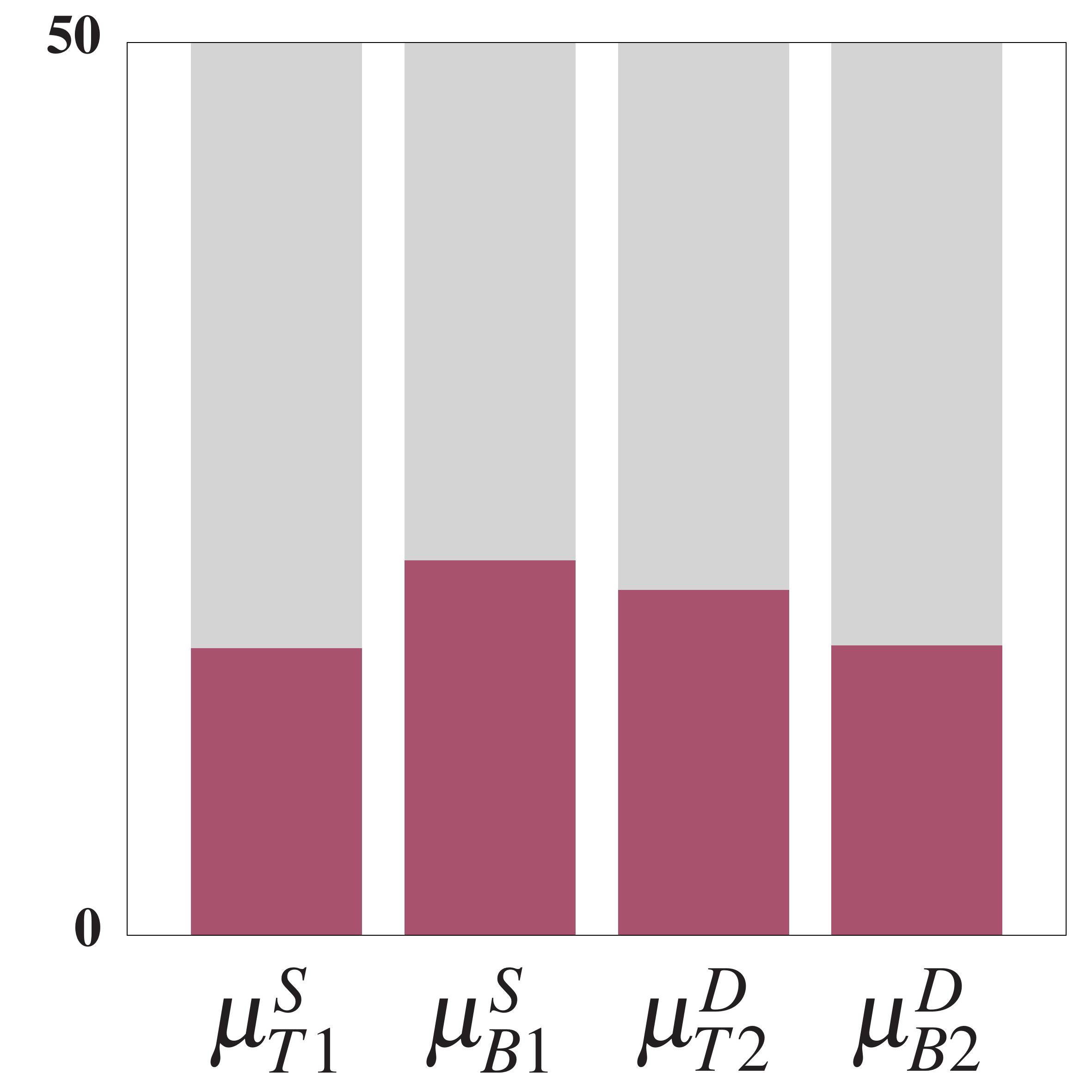}\label{fig:phi_mixParam}}\hfill~\\
\subfloat[\quad\quad(e)]{\includegraphics[width=0.48\columnwidth]{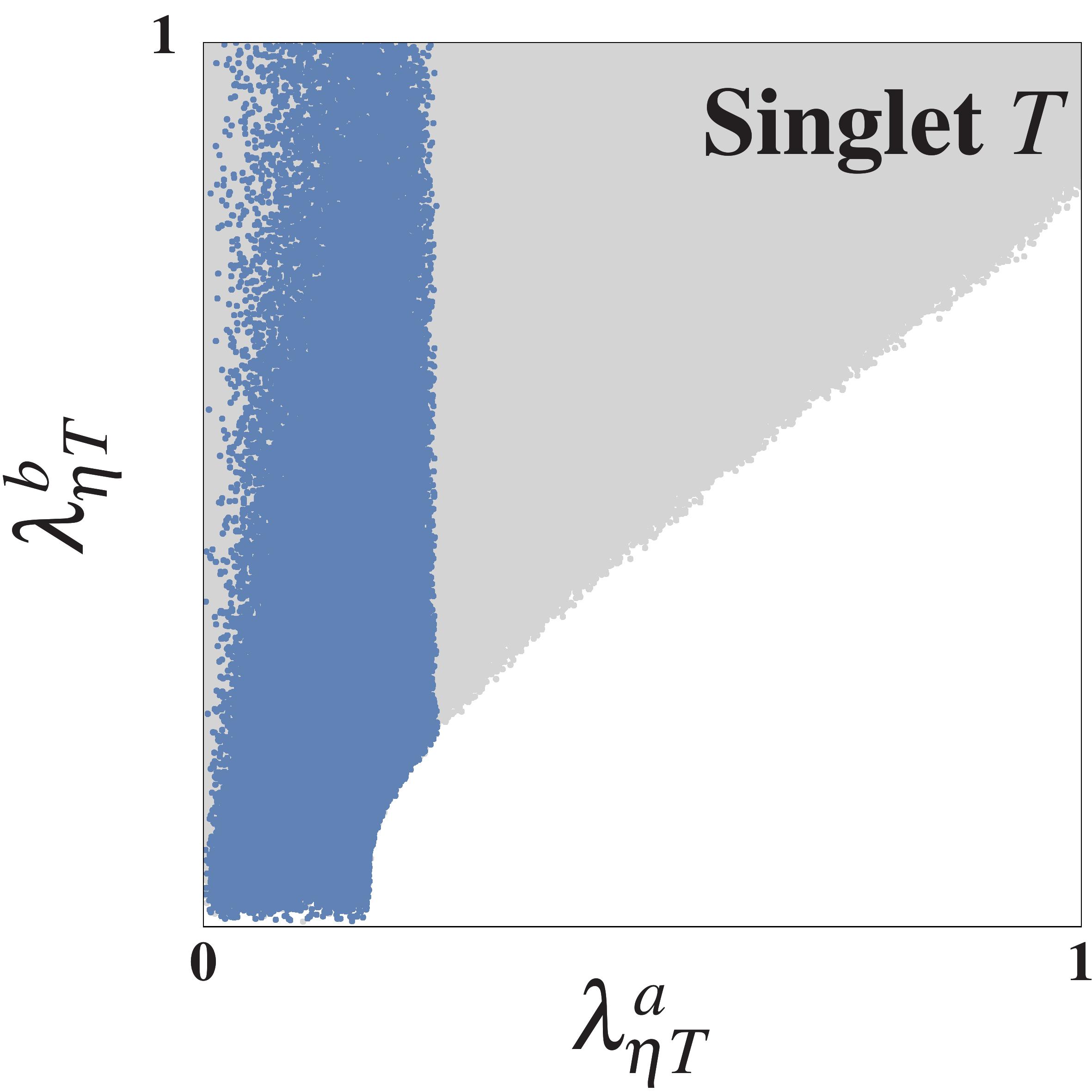}\label{fig:eta_SingTlm}}\hspace{0.2cm}
\subfloat[\quad\quad(f)]{\includegraphics[width=0.48\columnwidth]{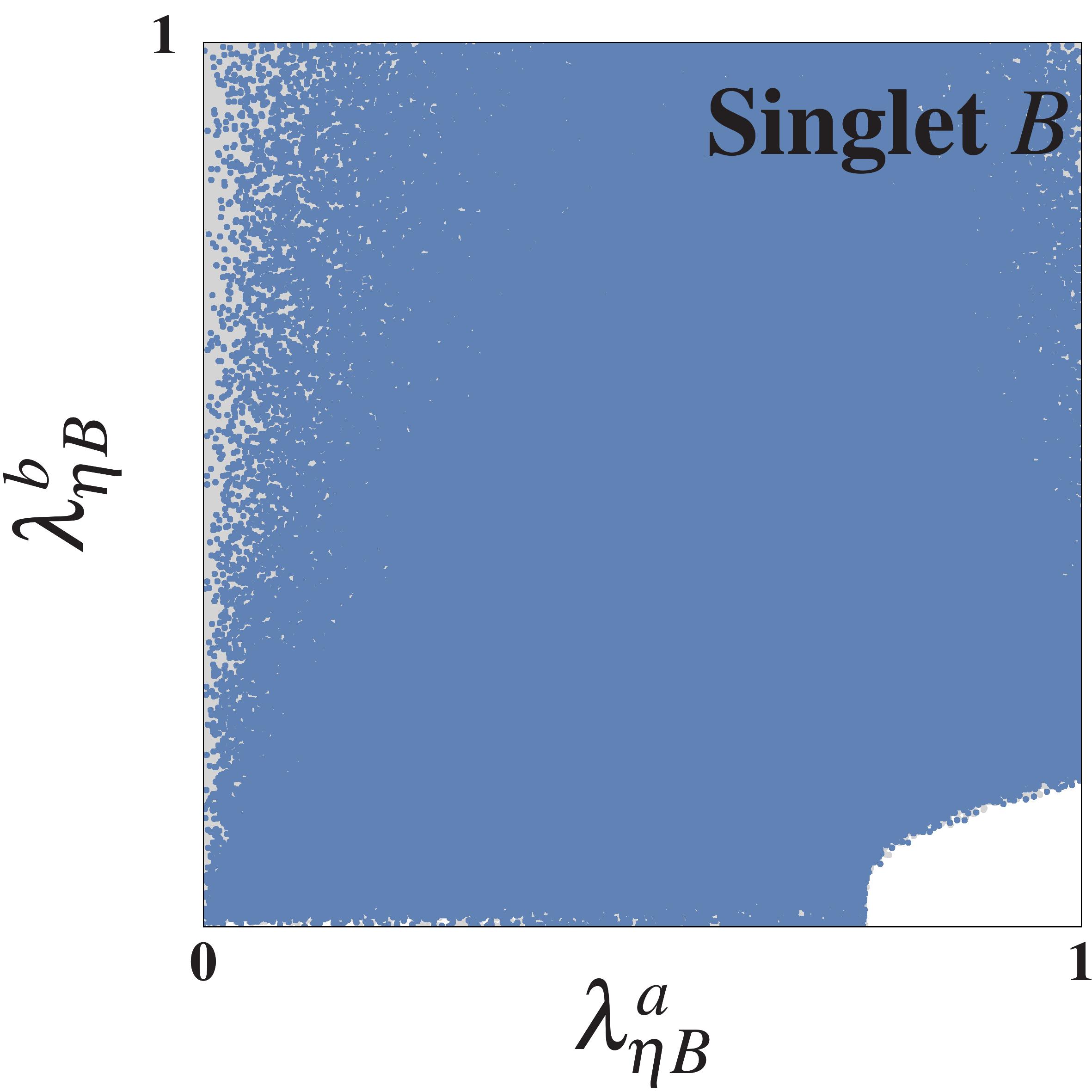}\label{fig:eta_SingBlm}}\hspace{0.2cm}
\subfloat[\quad\quad(g)]{\includegraphics[width=0.48\columnwidth]{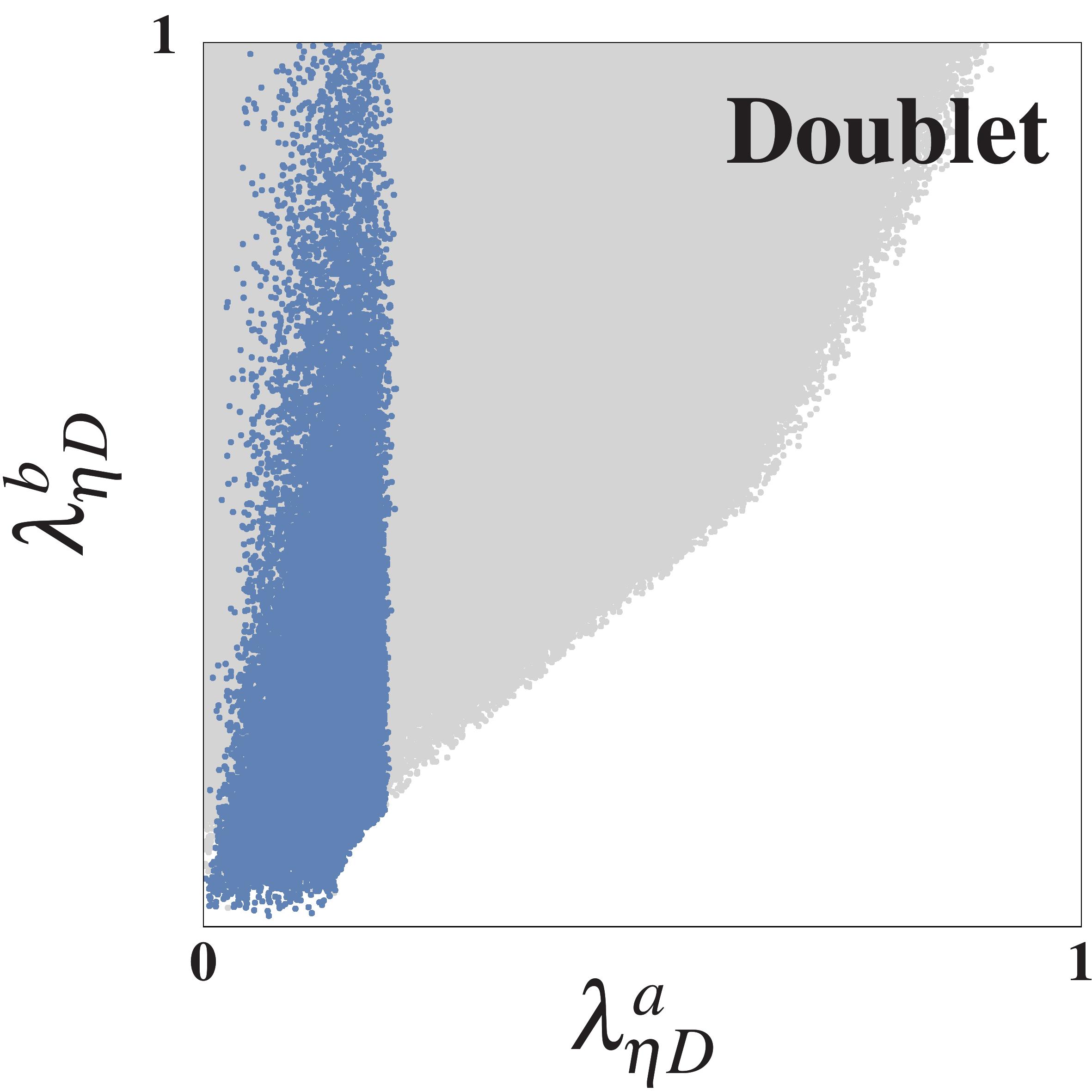}\label{fig:eta_Doublm}}\hspace{0.2cm}
\subfloat[\quad\quad(h)]{\includegraphics[width=0.48\columnwidth]{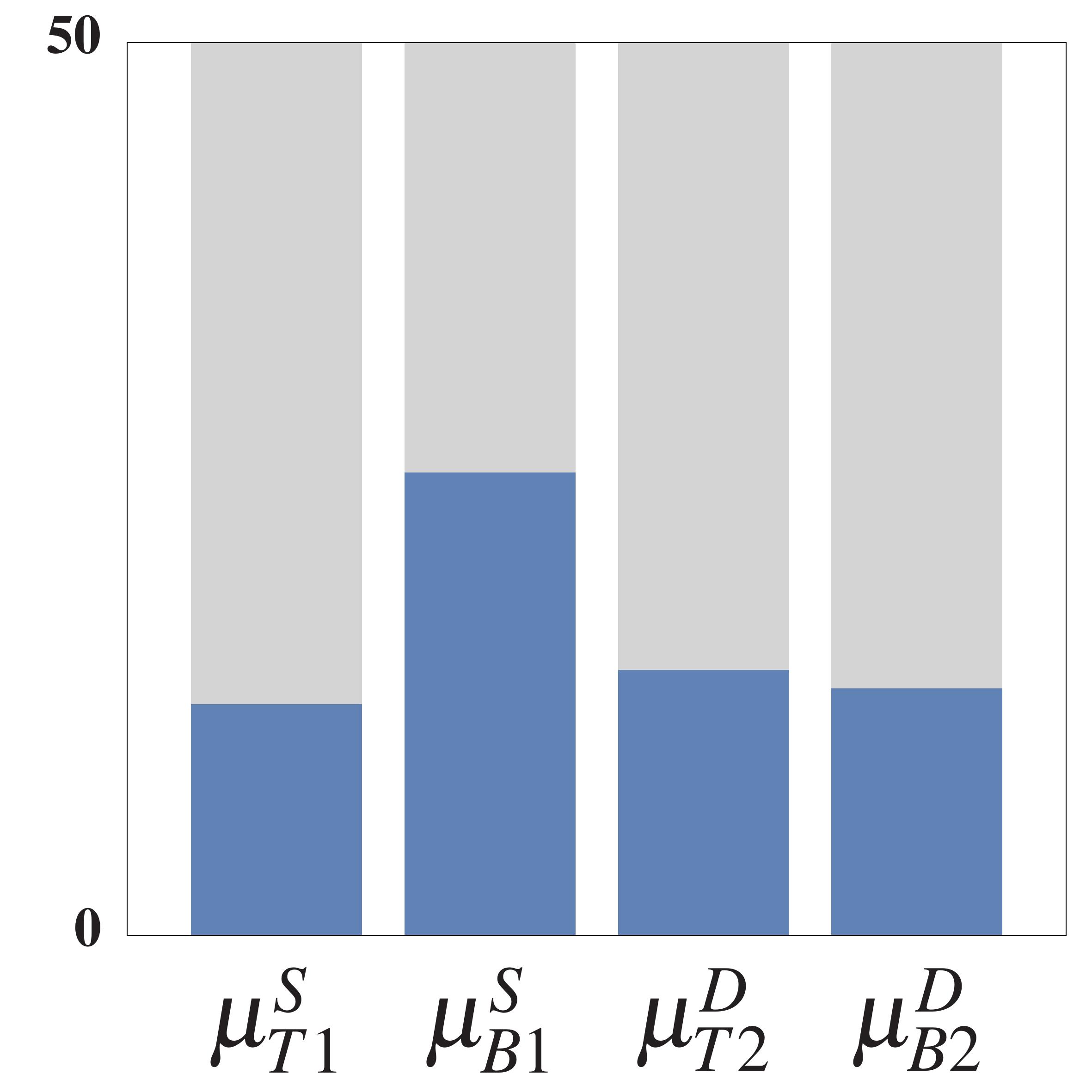}\label{fig:eta_mixParam}}\hfill~\\\label{fig:paramScans}
\caption{Results of numerical scans over the parameter spaces of the three models shown with projections. The bar graphs on the right [(d) and (h)] show the ranges  of the off-diagonal mass terms [see Eq.~\eqref{eq:massmatrix}] considered in the scans in GeV. The superscripts on the off-diagonal elements indicate the model---$S$ for Singlet and $D$ for Doublet. White regions in the plots are excluded by the constraints from Fig.~\ref{fig:boundsPhi}. The darker shades mark the regions where a loop decay of $\Phi$ dominates, i.e., $\bt^\Phi_{gg} > 0.5$.}
\end{figure*}
\subsection{Parameter scans}
\noindent
For the rest of section, we focus only on $\Phi=\phi$ since the pseudoscalar case is similar. To get an idea about the available parameter space surviving the bounds, we numerically scan over the model parameters incorporating the recast limits from Figs.~\ref{fig:boundsPhi} and \ref{fig:bounds_vlQ_rescaled} for a benchmark choice of $M_{F}= 1.2$ TeV and $M_\phi = 400 $ GeV. As mentioned above, we restrict the off-diagonal terms in the mass matrices, $\mu_{F_i}\lesssim 50$. This makes the mixing angles small, $\theta_L, \theta_R < 0.05$, which ensures the indirect bounds (like the correction to $\kp_{Z\bar b_Lb_L}$ etc. mentioned in the previous section) are respected.

Each of the singlet $T$ and $B$ models have three independent parameters of interest (one off-diagonal mass terms and two $\phi q_2q_1$ couplings), whereas the doublet model has four, since the $T$ and $B$ mass matrices share common elements (Eq.~\eqref{eq:massmatd}).  We use the following additional criteria for the scan:
\begin{enumerate}
    \item BR$(q_2 \to q_1\phi)$ should be greater than the rescaled experimental limits for $M_F=1.2$ TeV from Fig.~\ref{fig:bounds_vlQ_rescaled}. For example, for the Singlet $T$ model, BR$(t_2 \to t\phi) \gtrsim 70\%$.
    
    \item The effective coupling $\kp_{\phi gg}$ and $\bt^\phi_{\gm\gm}$ must satisfy the limit in Fig.~\ref{fig:boundsPhi}. So, for example,  $\kp^2_{\phi gg}\bt^\phi_{\gm\gm} \lesssim 2.7\times 10^{-9}$ for $M_\phi=400$ GeV.
    
    \item In addition, we mark the points that satisfy an additional criterion on the $\phi\to gg$ branching, $\bt^\phi_{g g}\geq50\%$.
\end{enumerate}
The last one is not a necessary condition but a choice. Its motivation differs from the first two. For $M_\phi>2m_q$, normally $\phi$ would significantly decay to a $q\bar q$ pair (a tree-level process). As a result, the pair production of $T$ can lead to the exotic $6t$ signature~\cite{Han:2018hcu}. However, as we shall see in the next section, there are other interesting and less explored signals of $\phi$ (like $\phi\to 2$-jets or $\gm\gm$) and the $T$ and $B$ quark. The third criterion takes us to the parameter regions where the loop-induced decays of the $\phi$ dominate. 

We show the results of the multidimensional scan for the singlet and doublet models with some projective plots in Fig.~\ref{fig:paramScans}. In these plots, all the grey points are allowed---clearly, there is no need to fine-tune the parameters to make the exotic decay mode dominant without violating the experimental bounds. The restriction from the LHC limit on $\kp_{\ph gg}\times \bt^\phi_{\gm\gm}$ for a $400$ GeV $\phi$ (Fig.~\ref{fig:boundsphi}) causes the empty areas on the bottom-right corners in the coupling plots (Figs.~\ref{fig:phi_SingTlm}, \ref{fig:phi_Doublm} and \ref{fig:eta_SingTlm}, \ref{fig:eta_SingBlm}, \ref{fig:eta_Doublm}). To understand this, we can consider, for example, the singlet $T$ model. We can write the diagonal couplings (that enter the $\phi\to\gm\gm$ loop, as shown in Eq.~\eqref{eq:S2gmgm}) of $\phi$ with the $t$ and $t_2$ quarks from Eq.~\eqref{eq:LagintS} as
\begin{align*}
    \lm_{\phi tt}:&\quad (\lm^b_{\phi T} s_Lc_R -\lm^a_{\phi T} s_Ls_R)\sim \lm^b_{\phi T}s_L,\\
    \lm_{\phi t_2t_2}:&\quad -(\lm^a_{\phi T} c_Lc_R +\lm^b_{\phi T} c_Ls_R)\sim -\lm^a_{\phi T}.
    \label{eq:SingModel_LoopProd}
\end{align*}
In the last step, we have ignored the relatively smaller terms suppressed by the $t-t_2$ mixing and set $c_{L/R}\sim1$. These imply that enhancing $\lm^a_{\phi T}$ ($\lm^b_{\phi T}$) increases the $t_2$ ($t$)-loop contribution to the $\phi\to\gm\gm$ decay and there is some cancellation between the two loops. Since $\phi$ dominantly couples to $t_2$, a large $\lm^a_{\phi T}$ is not favoured by the constraint on $\kappa_{\phi gg}^2\bt^\phi_{\gm\gm}$. However, because of the cancellation between the two quark contributions, it allows $\lm^a_{\phi T}$ and $\lm^b_{\phi T}$ to be large simultaneously. Similar arguments can be made in the other models as well. 
In the singlet $B$ model, the constraint on $\kp_{\phi gg}^2\bt^\phi_{\gm\gm}$ is weaker since the contribution of the loops with down-type quarks are suppressed than the up-type-quark loops by a factor of $(e_d/e_u)^4=1/16$.

We show the parameter regions where the loop-mediated $\phi$ decays dominate (i.e., the third criterion is satisfied) with darker shades. This essentially restricts the $\phi\to q_1q_1$ decay. Because of its mass, the top quark mixes easily with the $t_2$ quark than $b_1$ with $b_2$. Hence, this criterion restricts the parameter spaces in the singlet $T$ and the doublet models more than that in the singlet $B$ model. We can understand the behaviour of the parameters by looking at the singlet $T$ model once again. To reduce the $\phi tt $ coupling, the left-mixing angle $\theta_L$ should be small, restricting the off-diagonal mass element $\m_{T1}$ (which appears with $M_T$ in the numerator of Eq.~\eqref{eq:mixangleL}). However, there is another effect at play. As we enhance $\bt^\phi_{gg}$, we increase $\bt^\phi_{\gm\gm}$ as well (since they are proportional to each other, see Eqs.~\ref{eq:S2gmgm} and \ref{eq:S2gg}), and thus the limit on $\kp_{\phi gg}^2\bt^\phi_{\gm\gm}$ becomes more stringent, eliminating the region with large $\lm^a_\phi T$. Since the $\lm^b_{\phi T}$ term in the $\phi tt$ coupling is small due to the small left-mixing angle, the $\kp_{\phi gg}^2\bt^\phi_{\gm\gm}$ limit becomes insensitive to $\lm^b_{\phi T}$. A similar argument can be made for the doublet model as well. However, since the $\phi \to \gm\gm$ decay is small in the singlet $B$ model, demanding the gluon mode branching to be more than $50\%$ does not constrain the couplings further within the range we consider.

\begin{table}[t!]
     \centering
     \begin{tabular*}{\columnwidth}{ l@{\extracolsep{\fill}} p{0.36\columnwidth}  p{0.36\columnwidth} }
     \hline
     \multirowcell{2}{$q_2 \bar{q_2}$ \\decay} & \multicolumn{2}{c}{Possible final states}\\ \cline{2-3}
                                       & $q_2 = t_2$ & $q_2 = b_2$ \\
          \hline\hline
          \multirow{10}{*}{$q\Phi\;q\Phi$} & $2t + 4j$ & $2b + 4j$\\
                                          & $2t + 2\gamma + 2j$~\cite{Benbrik:2019zdp} & $2b + 2\gamma + 2j$ \\
                                          & $2t + 4\gamma$~\cite{Benbrik:2019zdp} & $2b + 4\gamma$ \\
                                          & $2t + 2b + 2j~({\#})$ & $2b + 2t + 2j~({\#})$ \\
                                          & $2t + 2b + 2\gamma~({\#})$ & $2b + 2t + 2\gamma~({\#})$ \\
                                          & $2t + 4b~({\#})$ & $2b + 4t~({\#})$ \\
                                          & $4t + 2j$ & $4b + 2j$ \\
                                          & $4t + 2\gamma$~\cite{Benbrik:2019zdp} & $4b + 2\gamma$ \\
                                          & $4t + 2b~({\#})$ & $4b + 2t~({\#})$ \\                     
                                          & $6t$~\cite{Han:2018hcu} & $6b$ \\
          \hline 
          \multirowcell{6}{$t\Phi\;bW$ \\ or \\ $b\Phi\;tW$} & $t + b + 4j$ & $t + b + 4j$ \\
                                                             & $t + b + 2\gamma + 2j$ & $t + b + 2\gamma + 2j$ \\
                                                             & $t + b + 2j + \ell + \slashed{E}$ & $t + b + 2j + \ell + \slashed{E}$ \\
                                                             & $t + b + 2\gamma + \ell + \slashed{E}$ & $t + b + 2\gamma + \ell + \slashed{E}$ \\
                                                             & $3t + b + 2j$ & $3b + t + 2j$ \\
                                                             & $3t + b + \ell + \slashed{E}$ & $3b + t + 2\gamma + \ell + \slashed{E}$ \\

          \hline
          \multirowcell{12}{$q\Phi\;q_1Z$ \\ or \\$q\Phi\;q_1h$} & $2t + 4j$ & $2b + 4j$ \\
                                                            & $2t + 4\gamma$ & $2b + 4\gamma$ \\
                                                            & $2t + 2b + 2j$ & $2b + 2j + 2\gamma$\\
                                                            & $2t + 2b + 2\gamma$ & $2b + 2j + 2\ell$\\
                                                            & $2t + 2j + 2\gamma$ & $2b + 2\ell + 2\gamma$ \\
                                                            & $2t + 2\ell + 2j$ & $2b + 2t + 2j~({\#})$ \\        
                                                            & $2t + 2\ell + 2\gamma$ & $4b + 2j$ \\
                                                            & $2t + 4b~({\#})$ & $4b + 2\gamma$ \\
                                                            & $4t + 2\gamma$ & $4b + 2\ell$\\
                                                            & $4t + 2b$ & $4b + 2t~({\#})$ \\
                                                            & $4t + 2j$ & $6b$\\
                                                            & $4t + 2\ell$ &  \\
                                                            
          \hline
     \end{tabular*}
     \caption{Possible pair production signatures when at least one heavy quark decays via the $q_2\to q_1\Phi$ mode. The signatures exclusive to the doublet model are indicated with a hash $(\#)$. We have ignored $\Phi\to Z\gm$ and the $\phi$ decays to heavy vector bosons as the corresponding modes are suppressed by the decays of the vector bosons.\label{tab:pp_processes}}
\end{table}
\begin{table*}
\centering{\linespread{3}
\begin{tabular*}{\textwidth}{l @{\extracolsep{\fill}} rrrrrrrrrrrl}
\hline
\multirowcell{2}[0pt][l]{BP} & $\mu_1$ & $\mu_{T2}$ & $\mu_{B2}$ & \multirowcell{2}[0pt][r]{$\lambda^a$} & \multirowcell{2}[0pt][r]{$\lambda^b$} & $M_\Phi$ & \multicolumn{3}{c}{$\Phi\to X$} & \multicolumn{2}{c}{$q_2 \to q\Phi$} & \multirowcell{2}[0pt][l]{Exotic $q_2$ decays}\\
\cline{8-10}\cline{11-12}
 &  (GeV) &  (GeV) &  (GeV) &  &  & (GeV) & $\beta^\Phi_{gg}$ & $\beta^\Phi_{bb}$ & $\beta^\Phi_{tt}$ & $\beta_{t\Phi}$ & $\beta_{b\Phi}$ & \\\hline\hline
\multicolumn{13}{c}{$q_2\to q_1\phi$} \\
\multicolumn{13}{l}{Singlet $T$} \\
 \hline
$T\phi1$ &  $9$ & --- & --- & $0.3$ & $0.5$ & $300$ &   $1$ & --- & --- & $1$ & --- & No $\phi \to tt$, $t_2\to t+\phi_{\rm boosted}$ \\
$T\phi2$ &  $9$ & --- & --- & $0.3$ & $0.2$ & $400$ & $0.5$ & --- & $0.5$ & $0.9$ & --- & $t_2\to ttt, t+\phi_{\rm boosted}$  \\
$T\phi3$ & $38$ & --- & --- & $0.5$ & $0.8$ & $400$ & $\sim0$ & --- &   $1$ & $0.8$ & --- & $t_2\to ttt$\\
$T\phi4$ & $11$ & --- & --- & $0.6$ & $0.2$ & $700$ & $0.6$ & --- & $0.4$ & $0.7$ & --- & Mostly $t_2\to t+2j$\\
$T\phi5$ & $16$ & --- & --- & $0.9$ & $0.7$ & $700$ & $0.2$ & --- & $0.8$ & $0.9$ & --- & Mostly $t_2\to ttt$  \\
\multicolumn{13}{l}{Singlet $B$} \\
\hline
$B\phi1^*$ & $14$ & --- & --- & $0.4$ & $0.1$ & $400$ & $0.5$ & $0.5$ & --- & --- & $0.4$ & \multirowcell{2}[0pt][l]{$b_2\to bbb, b+\phi_{\rm boosted}$} \\
$B\phi2^*$ & $26$ & --- & --- & $0.1$ & $0.2$ & $400$ & $\sim0$ & $1$ & --- & --- & $0.4$ & \\
$B\phi3$ &  $7$ & --- & --- & $0.7$ & $0.9$ & $700$ & $0.9$ & $0.1$ & --- & --- & $1$ & Mostly $b_2\to b+2j$\\
$B\phi4$ &  $8$ & --- & --- & $0.7$ & $0.9$ & $700$ & $0.8$ & $0.2$ & --- & --- & $1$ & Mostly $b_2\to bbb$\\
\multicolumn{13}{l}{Doublet} \\
\hline
$D\phi1$ & --- &  $2$ &  $4$ & $0.3$ & $0.2$ & $300$ & $0.7$ & $0.3$ & --- & $1$ & $1$ & \multirowcell{2}[0pt][l]{No $\phi \to tt$, $q_2\to q+\phi_{\rm boosted}$ } \\
$D\phi2$ & --- & $35$ & $25$ & $0.6$ & $0.6$ & $300$ &  ---  &  $1$  & --- & $0.9$ & $0.9$ & \\
$D\phi3$ & --- &  $4$ &  $4$ & $0.3$ & $0.2$ & $400$ & $0.9$ & $0.1$ & $\sim0$ & $1$ & $1$ & \multirowcell{2}[0pt][l]{$q_2\to q+\phi_{\rm boosted},q bb$ }\\
$D\phi4$ & --- & $13$ & $27$ & $0.1$ & $0.6$ & $400$ & $\sim0$ & $1$ &$\sim0$ & $1$ & $0.9$ & \\
$D\phi5$ & --- & $49$ & $2$ & $0.8$ & $0.9$ & $400$ & $0.1$ & $\sim0$ & $0.9$ & $0.9$ & $1$ & Mostly $q_2\to qtt$\\
$D\phi6$ & --- &  $4$ & $4$ & $0.4$ & $0.2$ & $700$ & $0.9$ & $\sim0$ & $\sim0$ & $1$ & $1$ & Mostly $q_2\to q+2j$ \\
$D\phi7$ & --- & $17$ & $45$ & $0.5$ & $0.8$ & $700$ & $\sim0$ & $0.9$ & $0.1$ & $1$ & $0.8$ & Mostly $q_2\to qbb$\\
$D\phi8$ & --- & $23$ &  $9$ & $0.2$ & $0.7$ & $700$ & $\sim0$ & $0.2$ & $0.8$ & $0.9$ & $1$ & Mostly $q_2\to qtt$\\
\hline\hline
\multicolumn{13}{c}{$q_2\to q_1\eta$} \\
\multicolumn{13}{l}{Singlet $T$} \\
 \hline
$T\eta1$ &  $\sim0$ & --- & --- & $0.2$ & $0.2$ & $300$ & $1.0$ & --- & ---  & $1$ & --- & No $\eta \to tt$, $t_2\to t+\eta_{\rm boosted}$ \\
$T\eta2$ &  $6$ & --- & --- & $0.3$ & $0.5$ & $400$ & $0.5$ & --- & $0.5$ & $1$ & --- & $t_2\to ttt, t+\eta_{\rm boosted}$  \\
$T\eta3$ & $50$ & --- & --- & $0.7$ & $0.9$ & $400$ & $\sim0$ & --- & $1$ & $0.8$ & --- & $t_2\to ttt$\\
$T\eta4$ &  $1$ & --- & --- & $0.2$ & $0.5$ & $700$ & $0.8$ & --- & $0.2$ & $1$ & --- & Mostly $t_2\to t+2j$ \\
$T\eta5$ & $35$ & --- & --- & $0.3$ & $0.8$ & $700$ & $\sim0$ & --- & $1$ & $0.8$ & --- & Mostly $t_2\to ttt$ \\
\multicolumn{13}{l}{Singlet $B$} \\
\hline
$B\eta1^*$ & $37$ & --- & --- & $0.9$ & $0.3$ & $400$ & $0.2$ & $0.8$ & --- & --- & $0.4$ & \multirowcell{2}[0pt][l]{$b_2\to bbb, b+\eta_{\rm boosted}$}\\
$B\eta2^*$ & $13$ & --- & --- & $0.5$ & $0.1$ & $400$ & $0.8$ & $0.2$ & --- & --- & $0.4$ & \\
$B\eta3$ & $4$ & --- & --- & $0.2$ & $0.1$ & $700$ & $1$ & $\sim0$ & --- & --- & $0.8$ & Mostly $b_2\to b+2j$ \\
$B\eta4$ & $30$ & --- & --- & $0.6$ & $0.4$ & $700$ & $0.1$ &  $0.8$ & --- & --- & $0.6$ & Mostly $b_2\to bbb$ \\
\multicolumn{13}{l}{Doublet} \\
\hline
$D\eta1$ & --- &  $9$ &  $2$ & $0.1$ & $0.2$ & $300$ & $0.9$ & $0.1$ & ---  & $0.9$ &   $1$ & \multirowcell{2}[0pt][l]{No $\eta \to tt$, $q_2\to q+\eta_{\rm boosted}$ }\\
$D\eta2$ & --- &  $5$ &  $7$ & $0.3$ & $0.7$ & $300$ & $0.1$ & $0.9$ & ---  & $1$ & $1$ & \\
$D\eta3$ & --- &  $1$ &  $6$ & $0.1$ & $0.1$ & $400$ & $0.8$ & $0.2$ & $\sim0$ & $1$ & $0.9$ & \multirowcell{2}[0pt][l]{$q_2\to q+\eta_{\rm boosted},q bb$ }\\
$D\eta4$ & --- & $21$ & $50$ & $0.4$ & $0.9$ & $400$ & $\sim0$ & $1$ & $\sim0$ & $1$ & $0.9$ &  \\
$D\eta5$ & --- & $46$ &  $4$ & $0.3$ & $0.8$ & $400$ & $0.1$ & $\sim0$ & $0.9$ & $0.9$ & $1$ & Mostly $q_2\to qtt$\\
$D\eta6$ & --- &  $4$ &  $8$ & $0.2$ & $0.1$ & $700$ & $0.9$ & $0.1$ & $\sim0$ & $0.9$ & $0.7$ & Mostly $q_2\to q+2j$ \\
$D\eta7$ & --- & $12$ & $21$ & $0.3$ & $0.5$ & $700$ & $0.2$ & $0.7$ & $0.1$ & $0.9$ & $0.8$ & Mostly $q_2\to qbb$\\
$D\eta8$ & --- & $39$ & $7$ & $0.6$ & $0.9$ & $700$ & $0.1$ & $\sim0$ & $0.9$ & $0.9$ &  $1$ & Mostly $q_2\to qtt$\\
\hline
\end{tabular*}}
\caption{For a representative choice of  $M_{q_2}=1.2$ TeV and three values of $M_{\Phi} = 300, 400, 700$ GeV, benchmark points with significant contribution to the exotic decay mode ($q_2\to q_1\Phi$) in the three models. The parameters are chosen such that $\mu_{1,2}< 50$ GeV and $\lm^{a,b}<1$. We also comment on the dominant topology of the exotic decay.
Benchmarks with non-negligible SM decays (i.e., $q_2\to q_1h,q_1Z,q_1^\prime W$) are marked with an asterisk ($^*$). }
\label{tab:BPs}
\end{table*}

\section{LHC signatures and a simple projection}\label{sec:projection}
\noindent
Adding the $q_2\to q_1\Phi$ decay mode of the heavy quarks leads to novel LHC phenomenology with several interesting signatures. For example, the pair production (which is essentially model-independent) of $t_2$ leads to the following possibilities:
\begin{align}
p p \to t_2 t_2 \to \left\{\begin{array}{lr}
t\Phi t\Phi &(\beta_{t\Phi}^2)\\
t\Phi bW &(2\beta_{t\Phi}\beta_{bW})\\
t\Phi tZ &\:(2\beta_{t\Phi}\beta_{tZ})\\
t\Phi tH &\:(2\beta_{t\Phi}\beta_{tZ})
\end{array}\right\}.
\end{align}
Here, we have shown the BR in each mode. Considering the decay modes of $\Phi$, we get a broad spectrum of final states. For example, let us consider the symmetric mode, i.e., the $q\Phi q\Phi$ mode and the fact that a $\Phi$ can decay to either a $gg$, $\gm\gm$, $tt$, or $bb$ pair. We can get the $6$-top signature (where both $\Phi$'s decay to $tt$ pairs)~\cite{Han:2018hcu}, or a final state with $4$ top quarks (only one $\Phi$ decays to a $tt$ pair) or $2$ top quarks. In the doublet model, one $\Phi$ can decay to a $tt$ pair while the other to a $bb$ pair leading to either a $4t+2b$ or $2t+4b$ final state. We list out the possibilities in Table~\ref{tab:pp_processes}.

\begin{figure*}
\captionsetup[subfigure]{labelformat=empty}
\centering
\subfloat[\quad\quad\quad(a)]{\includegraphics[width=0.9\columnwidth]{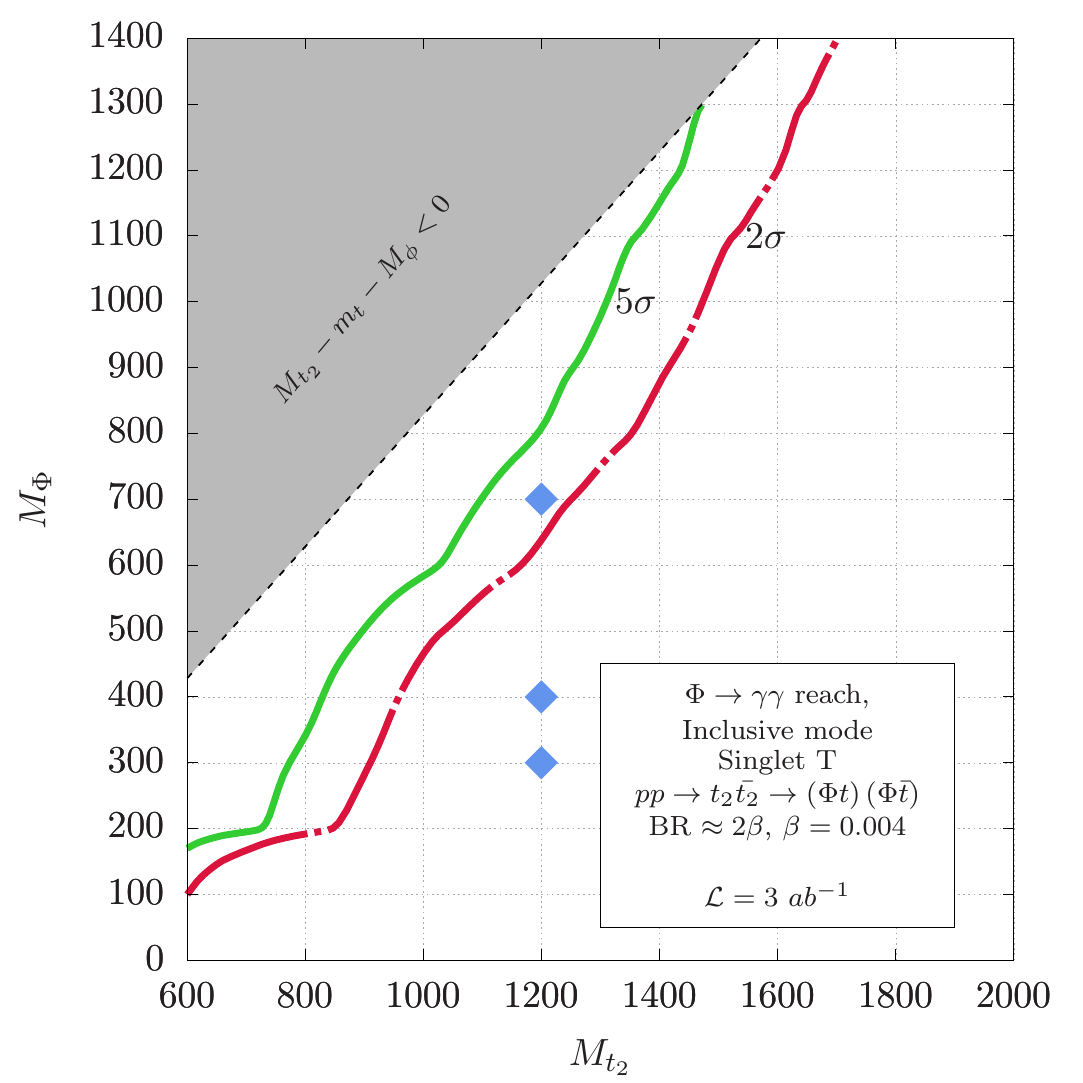}}\hspace{1cm}
\subfloat[\quad\quad\quad(b)]{\includegraphics[width=0.9\columnwidth]{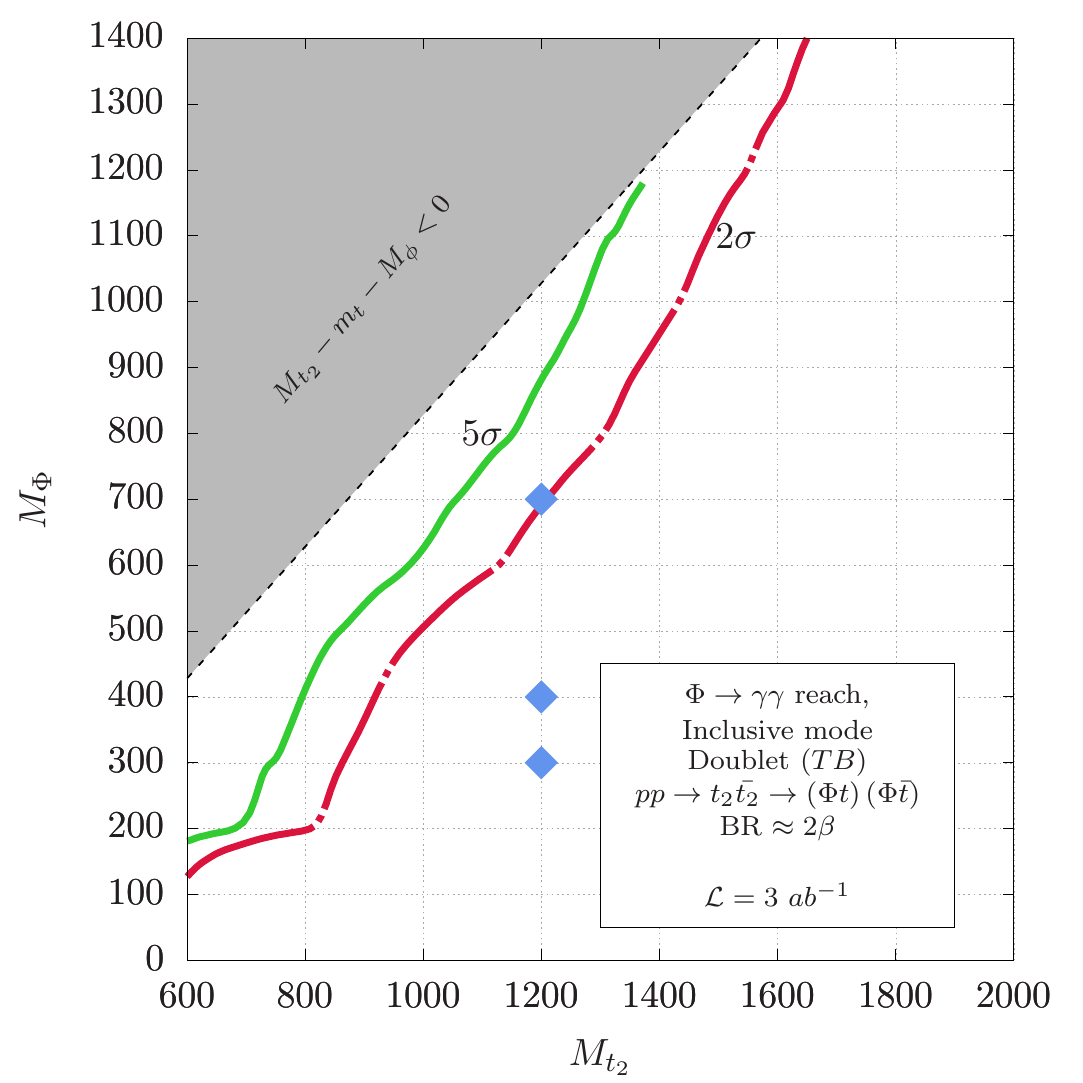}}
\caption{Reach at $\mathcal{L} = 3$ ab$^{-1}$ in the $pp\to t_2t_2\to tt+\gm\gm+X$ channel in the (a) singlet $T$ and (b) doublet models, assuming BR$(t_2 \to t\Phi) \approx 100\%$. The benchmark points    for $M_\Phi=300$, $400$, and $700$ GeV and $M_{t_2} = 1.2$ TeV (Table~(\ref{tab:BPs})) are shown as blue diamonds.}
\label{fig:proj_HLLHC}
\end{figure*}

We make some general observations below.
\begin{itemize}
\item In the doublet model, the $q_2\to q_1^\prime W$ decay is much suppressed than the $q_2\to q_1 Z/h$ decays (both have roughly equal BR). Hence, the $q\Phi q_1^\prime W$ modes (i.e., the ones with odd numbers of $t$ and $b$ quarks) are effectively exclusive to the singlet models and thus, can be useful to identify the weak representation of the heavy quarks. In the singlet models, when the VLQ is heavy, the BRs of $q_1^\prime W$, $q_1Z$, and $q_1h$ modes are in approximately $2:1:1$ ratio. (A few of the final states can arise from the conventional decays of the heavy quarks as well. For example, the $3b+t +$ jets or the $b+t +\ell+\slashed{E}_T+$ jets signatures can come from the $q_2q_2\to q_1h\,q_1^\prime W$ decay.)

\item The photons channels are cleaner than the hadronic ones~\cite{Benbrik:2019zdp}. However, these modes suffer from low BRs. The BR of the $\Phi\to \gm\gm$ decay is a few orders of magnitude smaller than  BR($\Phi\to gg$). For example, in the singlet models,
\begin{align}
    \label{eq:PhiGamGam_ratio}
    \frac{{\rm BR}(\Phi\to \gm\gm)}{{\rm BR}(\Phi\to gg)}= \frac{9\al^2}{2\al_s^2} Q_q^4,
\end{align}
where $Q_q$ is the electric charge of the heavy quark. This factor is about $0.004$ in the singlet $T$ model and about $0.0003$ in the singlet $B$ model. Hence the channels involving the $\Phi\to\gm\gm$ decay are negligible in the singlet $B$ model. 

\item
We do not consider the channels with the $\Phi\to Z\gm$ decay since $\bt^\Phi_{Z\gm}$ is small. While in the doublet model one gets $\bt^\Phi_{Z\gm}>\bt^\Phi_{\gm\gm}$,
the effective signal cross sections in these modes are reduced by the $Z$ decays. A similar argument is applicable for the $\phi\to VV$ decays. The $\Phi\to Z\gm$ decay mode is analysed in Ref.~\cite{Benbrik:2019zdp}.

\item
In a fully-hadronic analysis, one can use some kinematic features of the signal in different regions of the parameter space. In the models with a $T$ quark, if $M_{t_2}\gg M_\Phi$, both $\Phi$ and the top quark produced in the $t_2\to t\Phi$ decay would be boosted. Similarly, if  $M_\Phi\gg 2m_t$, the top quarks produced in the $\Phi\to tt$ decay would be boosted. The three-pronged nature of the boosted top quark(s) can efficiently enhance the signal over background ratio. 

\item Similarly, the final states produced in the standard decays of a TeV-range $q_2$ would be boosted (i.e., a boosted hadronically-decaying vector boson or a Higgs boson) and  give raise to two-prong fatjets.

\item A two-prong fatjet can also come from a $\Phi$ through the $\Phi\to gg, bb$ decays. As we have seen from the parameter scans, $\phi$ decays dominantly to two jets in significant parts of the parameter space in all models. In the singlet $B$ model, this is the signature of $\Phi$ in the entire parameter space. So far, a boosted two-prong $\Phi$-jet has not been used in any analysis of these models in the literature, even though it could be the dominant signal. We are currently analysing the prospects of identifying the signal with a boosted two-prong $\Phi$-jet. We shall present our results in a future publication. 
\end{itemize}

\subsection{Benchmark points}
\noindent
In Table~\ref{tab:BPs}, we show a representative set of parameters for $M_{q_2}=1.2$ TeV for an intuition about how the parameters relate to the various signal topologies described above. We consider three values of $M_\Phi$: $300$, $400$, and $700$ GeV. The first one is less than $2m_t$; hence, a $\Phi$ cannot decay to a $tt$ pair in this case, but it can in the second and third cases. In the second case, the $\Phi$ is boosted since $M_{q_2}\gg M_\Phi+m_{q_1}$, whereas in the third case, it will not be for $q_1=t$. The choice of the heavy quark mass is a representative one since it is clear from Fig.~\ref{fig:bounds_vlQ_rescaled} that $M_{q_2}\sim$ TeV is allowed if the BR$(q_2\to q_1\Phi)$ is not small. Hence the parameters are such that the $q_2\to q_1\Phi$ decay dominates, except the points in the singlet $B$ model marked with an asterisk ($^*$). The $b_2$ decays to the standard modes are non-negligible at these points (this is allowed by the relatively weaker limits from Fig.~\ref{fig:lims_T2} making the determination of the heavy quark representation relatively easier near these points). From Fig.~\ref{fig:bounds_vlQ_rescaled} we also see that as the masses of heavy-quarks increase, the lower limits on BR$(q_2\to q_1\Phi)$ relax. Hence, we can find similar parameter points with significant standard decays more easily.  In the table, we also point out the dominant signature of the heavy quark(s) for clarity. 

\subsection{Prospects at the HL-LHC}
\noindent
Estimating the optimal discovery/exclusion prospects for all the models is a nontrivial task, especially in the channels with the $\Phi\to gg$ decay. The results of our  study in these channels will be presented elsewhere. Instead, here we present a simple significance projection in the clean $t_2\to t\Phi \to t\gm\gm$ mode at the High Luminosity LHC (HL-LHC) as an illustration. We rely on the findings of Ref.~\cite{Benbrik:2019zdp} where the pair production of $t_2$ and their decay to $t\Phi \to t\gm\gm$ is considered. We use the cross sections for the background processes and the estimates for the signal and background selection efficiencies shown there to estimate the projected signal significance in the inclusive $pp\to t_2t_2\to tt+\gamma\gamma +  X$ channel.

We present our results in Fig.~\ref{fig:proj_HLLHC} where we plot the $2\sigma$ and $5\sigma$ contour lines. Since we are interested in the maximum reach, for every combination of $\{M_{t_2},M_\Phi\}$, we choose the parameters such that $\bt^\Phi_{\gamma\gamma}$ is maximum. For the singlet model,  it is easy to obtain $\bt^\Phi_{gg}\approx 1$. So we just take $\bt^\Phi_{\gamma\gamma}\approx 0.004$ following Eq.~\eqref{eq:PhiGamGam_ratio} in the entire $M_{t_2}-M_{\Phi}$ range we consider. 
Clearly, despite being the cleanest, because of the low branching, the $\Phi \to \gamma\gamma$ channel is not suitable to probe most of the parameter space.  However, as indicated by Table~\ref{tab:BPs},  there are other signatures that can be used (like the signals with a boosted $\Phi$-jet) to probe parts of this region.

So far, we have discussed only the pair production of the top partners. However, the top partners can also be produced singly. The single production channels could also lead to interesting new signatures in the presence of the singlet state. The cross sections of the single production channels would depend on the coupling between the VLQs with the SM particles.  Analysing the single productions would require different strategies (see, for example, Refs.~\cite{Mandal:2012rx,Mandal:2015vfa,Mandal:2016csb} for strategies to probe single productions of coloured particles) than the pair production searches but, depending on these couplings, single productions could be the dominant production mode of the top partners.


\section{Summary and conclusions}\label{sec:conclusion}

\noindent
Exhaustive searches for VLQs in the standard channels, where they decay to the SM fields, and non-observation of any deviation from the SM predictions at the LHC motivate us to look for them in new decay channels. In this paper, we have charted out the possibility of exploring heavy vectorlike top and bottom partners decaying to a new weak-singlet colourless scalar or pseudoscalar and a third-generation quark. As motivated in the introduction, such possibilities can arise in many new physics models. Therefore, exploring these new decays of the VLQs in the upcoming run-$3$ of the LHC would be of prime importance. We have considered simple phenomenological models covering the possible weak representations of the VLQs that can couple with a SM-singlet pseudo(scalar). We have reinterpreted the latest mass-exclusion limits for $T$ and $B$ quarks in terms of the BRs in the new decay modes. With the increasing branching in the extra decay mode, the existing limits on VLQs can relax by up to $\sim 300-500$ GeV. Beyond their weak representation, the recast limits are independent of the exact nature of the additional decay modes; hence they are applicable in a broader range of models than those considered here. 

Incorporating the reinterpreted direct-search limits on VLQs and the (pseudo)scalar, we found that the parameter space is wide open and does not need any fine-tuning or abnormally large mixing with new quarks or large off-diagonal couplings. Hence, on the theoretical side, the next-to-minimal avatars of most well-motivated models featuring VLQs and a (pseudo)scalar can be easily mapped to the open parameter space. For example, the off-diagonal couplings involving third-generation quarks and their vectorlike partners tend to be small in the warped extra-dimension models (see, e.g., Ref.~\cite{Gopalakrishna:2013hua}). So, following our results, one can consider such a set-up with an extra spinless field (e.g., Ref.~\cite{Gopalakrishna:2015wwa}) without conflicting with the LHC data. On the experimental side, we have charted out a host of interesting and unexplored collider signatures. We have presented a set of benchmark points to probe different signatures as a guideline for future VLQ searches at the LHC. We have also performed a simple projection study in the clean $\Phi\to\gm\gm$ channel and indicated how other channels could be used to probe additional regions.

We point out one  channel of particular interest where a VLQ decays to a pseudo(scalar), which decays to two gluons. The di-gluon mode is the dominant decay of the singlet if the tree-level decays are kinematically forbidden. However, even if the tree-level decays are allowed, the loop-induced decay to a pair of gluons can still dominate over the tree-level ones---our random scans have shown that this is the case over a large region of the available parameter space of every model. Hence, in those regions, it can act as the discovery channel (we will report the HL-LHC prospects of this channel in a forthcoming paper).

There are some cases where our results would not be directly applicable. For example, we have assumed that before EWSB, $\Phi$ does not couple exclusively with the SM fields. However, one can think of models where $\Phi$ couples with the Higgs field. In that case, one might need to consider additional decays and the corresponding experimental bounds. Similarly, one can consider models with more than one singlet/doublet VLQs or other heavy fields that couple with the VLQs and $\Phi$. One can easily follow our prescription to obtain the available parameter space in all such cases.

\bc{\bf ACKNOWLEDGEMENTS}\ec
\noindent
We thank the anonymous referee for pointing out the presence of a redundant parameter in our analysis. We thank Arvind Bhaskar for reading and commenting on the manuscript. A. B. is supported by the STFC under Grant No. ST/T$000945$/$1$. T. M. thanks the SHIFT project members for initial discussions and acknowledges the use of the high-performance computing facility at IISER-TVM. S. M. acknowledges support from the Science and Engineering Research Board, India, under Grant No. ECR/$2017$/$000517$. C. N. acknowledges DST-Inspire for his fellowship. 

\def\bibfont{\small}
\bibliography{VLQ}{}
\bibliographystyle{JHEPCust}

\end{document}